\documentclass[11pt]{article}

\usepackage[final]{acl}

\usepackage{times}
\usepackage{latexsym}
\usepackage{subcaption}
\usepackage{algorithm}
\usepackage{algpseudocode}
\usepackage{amsmath}
\usepackage{setspace}
\usepackage[T1]{fontenc}

\usepackage[utf8]{inputenc}

\usepackage{microtype}

\usepackage{inconsolata}

\usepackage{graphicx}

\usepackage[utf8]{inputenc} 
\usepackage[T1]{fontenc}    
\usepackage{hyperref}       
\usepackage{url}            
\usepackage{booktabs}       
\usepackage{amsfonts}       
\usepackage{nicefrac}       
\usepackage{microtype}      
\usepackage[table,dvipsnames]{xcolor}
\usepackage{pifont}
\usepackage{amssymb}
\usepackage{bbding}
\usepackage{bm}
\usepackage{amsmath}
\usepackage{xcolor}
\usepackage{ragged2e}

\usepackage{graphicx}

\usepackage{algorithm}
\usepackage{algpseudocode}

\usepackage{tikz}
\usetikzlibrary{positioning, shapes.geometric, arrows.meta, shadows, calc}

\definecolor{heat1}{RGB}{220, 240, 255} 
\definecolor{heat2}{RGB}{160, 210, 250} 
\definecolor{heat3}{RGB}{60, 120, 220}  
\definecolor{heat4}{RGB}{10, 50, 140}   

\definecolor{maskgray}{RGB}{200, 200, 200} 

\definecolor{tokengreen}{RGB}{220, 255, 220}

\DeclareFixedFont{\ttb}{T1}{txtt}{bx}{n}{9} 
\DeclareFixedFont{\ttm}{T1}{txtt}{m}{n}{9}  
\usepackage{listings}
\definecolor{deepblue}{rgb}{0,0,0.8}
\definecolor{deepred}{rgb}{0.6,0,0}
\definecolor{deepgreen}{rgb}{0,0.5,0}

\usepackage{pythonhighlight}
\usepackage{enumitem}
\usepackage{wrapfig}
\usepackage{multirow}
\usepackage{graphicx}
\usepackage{tcolorbox}
\usepackage{colortbl}
\usepackage{makecell}
\definecolor{light-gray}{gray}{0.92}
\newcommand{\xhdr}[1]{\noindent{{\bf #1.}}}

\usepackage[table]{xcolor}
\usepackage{xcolor}
\usepackage{pifont}
\usepackage{arydshln}
\newcommand{\cmark}{\textcolor{green!60!black}{\ding{51}}}
\newcommand{\xmark}{\textcolor{red!70!black}{\ding{55}}}
\newcommand{\smark}{\textcolor{green!60!black}{\ding{116}}}

\newcommand\blfootnote[1]{%
  \begingroup
  \renewcommand\thefootnote{}\footnote{#1}%
  \addtocounter{footnote}{-1}%
  \endgroup
}

\usepackage{xurl}

\title{MRMAD: A Multi-Round Multi-Audio Benchmark for Evaluating Acoustic Degradation Perception in Large Audio-Language Models}

\author{
    \textbf{Yize Li\textsuperscript{1, 2$*$$\dagger$}}\quad
    \textbf{Ningyuan Yang\textsuperscript{3$*$}}\quad
    \textbf{Sile Yin\textsuperscript{2}}\quad
    \textbf{Sindhuja Thogarrati\textsuperscript{2}}\quad
\\
    \textbf{Sung-En Chang\textsuperscript{2}}\quad
    \textbf{Andrew C. Singer\textsuperscript{3}}\quad
    \textbf{Xue Lin\textsuperscript{1}}\quad
    \textbf{Chuan-Che Huang\textsuperscript{2}}\quad
    \textbf{Shuo Zhang\textsuperscript{2$\dagger$}}
\\
    \textsuperscript{1}Northeastern University\quad
    \textsuperscript{2}Bose Corporation\quad
    \textsuperscript{3}Stony Brook University
\\
}

\begin{document}
\maketitle

\blfootnote{%
\textsuperscript{*}Equal contribution. Work done during an internship at Bose Corporation.
\textsuperscript{$\dagger$}Correspondence to: Yize Li~<\href{mailto:li.yize@northeastern.edu}{\nolinkurl{li.yize@northeastern.edu}}>,
Shuo Zhang~<\href{mailto:SHUO_ZHANG@bose.com}{\nolinkurl{SHUO_ZHANG@bose.com}}>.
}

\begin{abstract}
Large audio-language models (LALMs) have shown promising progress in understanding speech, music, and general sound events, yet their ability to reason about how audio signals are degraded remains underexplored. Existing benchmarks primarily evaluate semantic understanding, event recognition, or high-level audio reasoning, leaving a  basic question unanswered: 
\textbf{Do LALMs understand the differences in audio quality?} 
We introduce MRMAD, a Multi-Round Multi-Audio Degradation benchmark for evaluating audio degradation perception and understanding in LALMs. MRMAD spans speech, music, and sound, and frames evaluation as multi-turn dialogues across multiple audio inputs, requiring models to identify types of degradation, compare severity, and perceive corruption changes across turns. Unlike current single-turn audio-language benchmarks, MRMAD evaluates whether LALMs can maintain consistent degradation hypotheses with new evidence and comprehend low-level acoustic phenomena over multi-turn dialogues. Through a systematic evaluation of 18 representative LALMs from non-thinking to reasoning and Omni models, we find that current models often recognize coarse content while failing to diagnose, compare, or reason about degradations reliably. Human evaluations further reveal a significant perception gap between LALMs and human listeners. MRMAD thus exposes a critical yet overlooked aspect of audio-language understanding and provides a diagnostic foundation for building future LALMs that are robust to real-world acoustic conditions\footnote{Code and benchmark: \href{https://github.com/Bose/MRMAD}{\textcolor{blue}{https://github.com/Bose/MRMAD}}}.
\end{abstract}

\section{Introduction}
\label{sec:intro}
\begin{figure*}[t]
\centering
\includegraphics[width=1\textwidth]{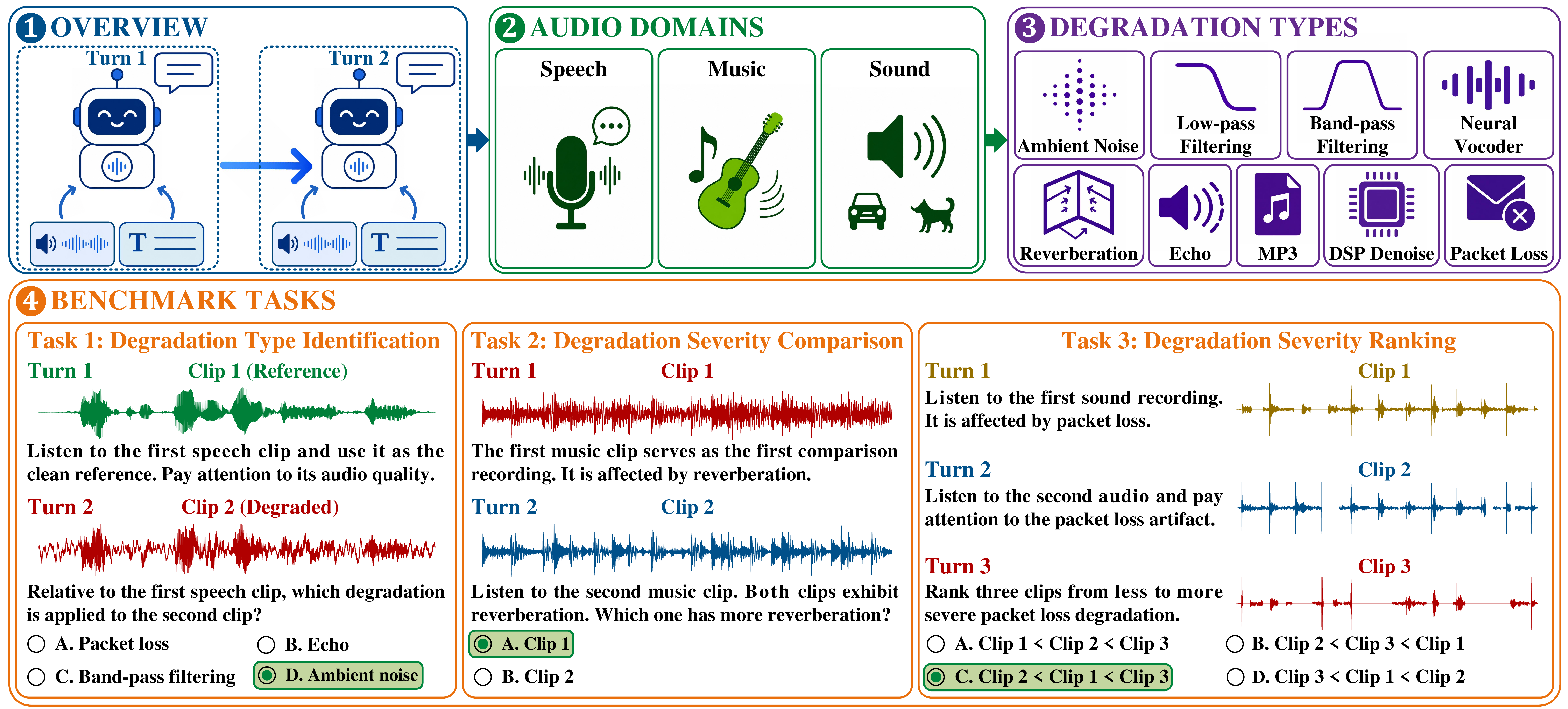}
\caption{
MRMAD benchmark overview. 
MRMAD evaluates multi-turn multi-audio degradation perception and quality understanding across three audio domains, nine degradation types, and three tasks.
}
\label{fig:benchmark}
\end{figure*}

Large Audio-Language Models (LALMs) and Omni language models (OLMs) have recently shown strong progress in audio-centered interaction, enabling models to perceive and reason over speech, music, and sound events through natural language interfaces~\citep{tang2023salmonn,chu2024qwen2audio,gemini2024gemini15,kong2025,li2025preference,wang2025reasoning,you2026gamma,li11541222,xiao2026adapting,sun2026act,tong2026autagent}. 
Along with these models, numerous audio-language benchmarks have been proposed to evaluate audio understanding, instruction following, and complex reasoning across diverse auditory domains~\citep{weck2024muchomusic,yang2024air,wang2025audiobench,sakshi2025mmau,ma2025mmar,wang2025mmsu,sun2026sonicbench,yang2026mugen}. 
These benchmarks have substantially advanced the evaluation of semantic audio understanding, including speech comprehension, sound event recognition, music reasoning, and audio-grounded question answering.

Despite this progress, a fundamental auditory capability remains underexplored---judging audio quality, both objective and perceptual.
Real-world audio recordings are rarely pristine and can be affected by diverse degradations, including background noise, reverberation, codec compression, bandwidth limitation, and temporal discontinuities~\cite{pascual2017segan,lemercier2025diffusion,yang2026survey}.
For humans, recognizing those distortions is often a prerequisite for assessing audio quality and determining how audio should be restored or enhanced.
For LALMs, however, it remains unclear whether they can reliably perceive these low-level quality differences rather than relying primarily on high-level semantic cues.

Existing benchmarks are not designed to answer this question. 
General audio-language benchmarks such as AIR-Bench~\citep{yang2024air}, AudioBench~\citep{wang2025audiobench}, MMAU~\citep{sakshi2025mmau}, MMSU~\citep{wang2025mmsu}, and MMAR~\citep{ma2025mmar} mainly evaluate content understanding, instruction following, domain knowledge, or multi-step reasoning. 
Music-specific benchmarks, including MuChoMusic~\citep{weck2024muchomusic}, focus on musical concepts and cultural or functional reasoning. 
Speech quality benchmarks such as QualiSpeech~\citep{wang2025qualispeech} move closer to low-level perceptual assessment, but remain limited to speech and single-audio evaluation.

To systematically evaluate audio quality understanding, an intrusive setting with reference audio is needed for comparison.
However, concatenating multiple audio clips into a single input would require models to perform accurate temporal localization to determine which segment corresponds to each prompt, introducing an additional confounding factor.
A multi-round, multi-audio format is therefore better suited for this setting: each round provides one audio clip with its corresponding text prompt, enabling comparison through conversational context.

To address this gap, we introduce \textbf{\textit{MRMAD}}, a \textbf{M}ulti-\textbf{R}ound \textbf{M}ulti-\textbf{A}udio \textbf{D}egradation benchmark for evaluating audio quality understanding in LALMs.
MRMAD comprises three basic tasks---\textit{degradation type identification} (DTI), \textit{degradation severity comparison} (DSC), and \textit{degradation severity ranking} (DSR)---spanning across speech, music, and sound, while covering nine common degradation types.
To the best of our knowledge, it is the \textbf{\textit{first}} unified benchmark that explicitly targets audio quality perception and understanding in LALMs.

Our contributions are summarized as follows:
\begin{itemize}
    \item We introduce MRMAD, the first multi-round, multi-audio benchmark for evaluating audio quality understanding in LALMs, complementing evaluations that focus on semantic audio understanding and general reasoning.
    \item We design three tasks comprising 8,400 multiple-choice questions across speech, music, and sound to assess whether models can identify degradation types, compare degradation severity, and rank degraded clips.
    \item We comprehensively evaluate 18 LALMs on MRMAD using a reproducible evaluation framework, revealing substantial limitations of current models in degradation-aware auditory perception, cross-audio comparison, and multi-turn severity reasoning.
    \item We conduct a small-scale evaluation comparing representative LALMs with human listeners, demonstrating that LALMs still notably underperform humans on our benchmark.
\end{itemize}
\section{Related Work}
\label{sec:relate}

\begin{table*}[t]
\centering
\small
\setlength{\tabcolsep}{5pt}

\caption{Comparison between MRMAD and existing audio-language benchmarks in terms of benchmark size (number of questions), audio-domain coverage, multi-turn design, and evaluation focus.}
\label{tab:benchmark_comparison}
{
\renewcommand{\arraystretch}{0.95}
\begin{tabular}{lcccccc}
\toprule
\multirow{2}{*}[-0.6ex]{\textbf{Benchmark}}
& \multirow{2}{*}[-0.6ex]{\textbf{Size}}
& \multicolumn{3}{c}{\textbf{Domain}}
& \multirow{2}{*}[-0.6ex]{\textbf{Multi-Turn}}
& \multirow{2}{*}[-0.4ex]{\makecell{\textbf{Degradation Perception}\\\textbf{\& Quality Understanding}}} \\
\cmidrule(lr){3-5}
& & \textbf{Speech} & \textbf{Music} & \textbf{Sound} & & \\
\midrule
AIR-Bench~\cite{yang2024air} & 19k & \cmark & \cmark & \cmark & \xmark & \xmark \\
MuChoMusic~\cite{weck2024muchomusic} & 1.2k & \xmark & \cmark & \xmark & \xmark & \xmark \\
AudioBench~\cite{wang2025audiobench} & 100k+ & \cmark & \cmark & \cmark & \xmark & \xmark \\
QualiSpeech~\cite{wang2025qualispeech} & 4.08k & \cmark & \xmark & \xmark & \xmark & \cmark \\
MMAU~\cite{sakshi2025mmau} & 10k & \cmark & \cmark & \cmark & \xmark & \xmark \\
MMAR~\cite{ma2025mmar} & 1k & \cmark & \cmark & \cmark & \xmark & \smark \ (Extremely Limited) \\
MMSU~\cite{wang2025mmsu} & 5k & \cmark & \xmark & \xmark & \xmark & \xmark \\
\rowcolor{gray!30}
\textbf{MRMAD} (\textit{ours}) & 8.4k & \cmark & \cmark & \cmark & \cmark & \cmark \\
\bottomrule
\end{tabular}
}
\end{table*}

\xhdr{Large Audio-Language Models}\
Recent LALMs have extended unimodal LLMs to the auditory modality by connecting audio encoders with instruction-tuned language models. Models such as SALMONN~\citep{tang2023salmonn} and Qwen2-Audio~\citep{chu2024qwen2audio} demonstrate broad capabilities across speech, music, and general sound.  
Large Audio Reasoning Models (LARMs) further emphasize complex auditory reasoning, long-audio understanding, and chain-of-thought supervision~\citep{ghosh2024gama,xie2025audioreasoner,ghosh2025audioflamingo2}. In parallel, OLMs such as Qwen-Omni~\citep{xu2025qwen25omni,xu2025qwen3omni} and Gemini~\citep{gemini2024gemini15,geminiteam2025}, move toward unified interaction across text, audio, and vision. Despite these advances, existing models are primarily optimized and evaluated for semantic understanding, instruction following, and high-level reasoning. Their ability to perceive and reason about low-level audio quality attributes remains insufficiently studied, especially in multi-turn, multi-audio conversational settings~\citep{yang2026mugen}.

\xhdr{Audio-Language Benchmarks}\
A growing number of benchmarks have been proposed for audio-language models. AIR-Bench~\citep{yang2024air} evaluates LALMs through chat-based tasks over diverse audio domains, while AudioBench~\citep{wang2025audiobench} covers speech understanding, audio scene understanding, and paralinguistic voice understanding. In addition, MMAU~\citep{sakshi2025mmau} increases difficulty with expert-level knowledge and multi-step reasoning, and MMAR~\citep{ma2025mmar} extends evaluation to multi-domain reasoning tasks.
MMSU~\citep{wang2025mmsu} focuses on fine-grained spoken language understanding and reasoning, including linguistic and paralinguistic phenomena. For music, MuChoMusic~\citep{weck2024muchomusic} evaluates musical concepts, cultural context, and reasoning over music tracks. Closest to our focus, QualiSpeech~\cite{wang2025qualispeech} investigates speech quality assessment with language descriptions and reasoning. However, these benchmarks primarily focus on content understanding, event inference, domain-specific knowledge, or general reasoning, rather than systematically evaluating audio quality understanding in multi-turn, multi-audio settings.
Table~\ref{tab:benchmark_comparison} compares our MRMAD with recent audio-language benchmarks.

\section{MRMAD Benchmark
}
\label{sec:dataset}
\subsection{Overview}
MRMAD assesses whether LALMs can identify, compare, and rank common audio degradations across diverse audio domains.
The benchmark consists of 8,400 multiple-choice questions paired with carefully constructed audio clips.
Spanning speech, music, and sound, MRMAD covers nine degradation types: \textit{ambient noise}, \textit{low-pass filtering}, \textit{band-pass filtering}, \textit{MP3 compression}, \textit{reverberation}, \textit{echo}, \textit{neural vocoder artifacts}, \textit{DSP denoising artifacts}, and \textit{packet loss}.

\begin{figure*}[t]
\centering
\includegraphics[width=1\textwidth]{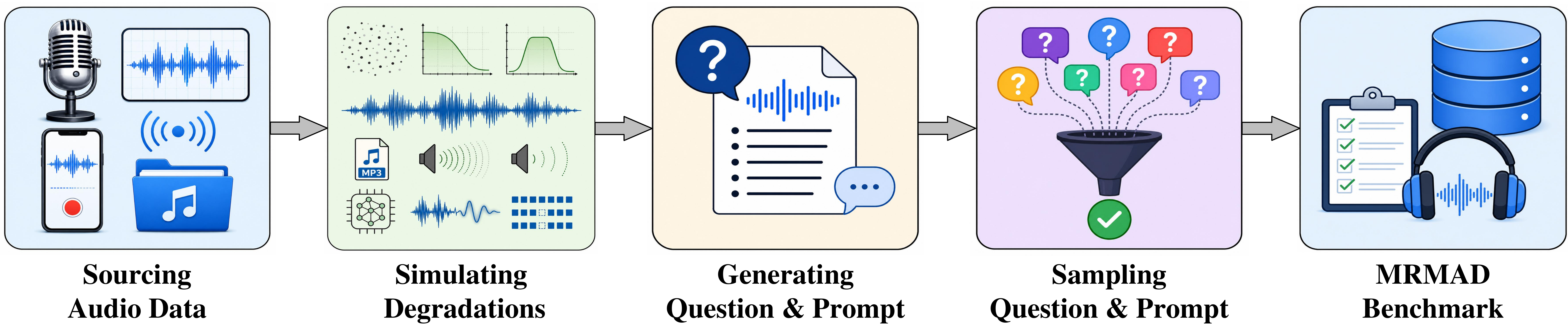}
\caption{
Data construction pipeline of MRMAD.
For each collected audio clip, we simulate all domain-specific degradations at three severity levels.
We then generate task-specific multi-turn questions and prompts for DTI, DSC, and DSR based on the degraded audio clips.
Finally, we sample a balanced subset to form the benchmark.
}
\label{fig:pipeline}
\end{figure*}

MRMAD contains three tasks:
\textit{degradation type identification} (DTI), \textit{degradation severity comparison} (DSC), and \textit{degradation severity ranking} (DSR).
In DTI, models identify the degradation type in a degraded clip with access to its clean reference.
In DSC, models compare two degraded clips and determine which one has higher or lower severity.
In DSR, models rank three degraded clips by degradation severity.
These tasks use a multi-turn, multi-audio format, requiring models to integrate information across rounds rather than making isolated single-clip predictions. 
This design enables a more comprehensive evaluation of fine-grained perceptual reasoning over audio quality. Table~\ref{tab:dataset_source} summarizes the source datasets used to construct MRMAD across speech, music, and sound. Figure~\ref{fig:pipeline} illustrates the overall data construction pipeline.

\begin{table}[htbp]
\centering
\small
\setlength{\tabcolsep}{4.90pt}
\caption{Summary of the benchmark composition across speech, music, and sound. 
For each audio type, we report the dataset sources and the number of questions.}
\label{tab:dataset_source}
\begin{tabular}{ccc}
\toprule
\textbf{Type}
& \textbf{Dataset Sources}
& \textbf{Num} \\
\midrule
\multirow{4}{*}{Speech}
& VCTK~\cite{yamagishi2012english}
& \multirow{4}{*}{3,150} \\
& LJSpeech~\cite{ljspeech17} & \\
& HiFi-TTS~\cite{bakhturina2021hi} & \\
& LibriSpeech~\cite{panayotov2015librispeech} & \\

\midrule
\multirow{3}{*}{Music}
& URMP~\citep{li2018creating}
& \multirow{3}{*}{2,800} \\
& MUSDB18-HQ~\cite{MUSDB18HQ} & \\
& FMA-small~\cite{fma_dataset} & \\

\midrule
\multirow{2}{*}{Sound}
& FSD50K~\cite{fonseca2021fsd50k}
& \multirow{2}{*}{2,450} \\
& TAU Urban 2019~\cite{mesaros2019tau} & \\

\bottomrule
\end{tabular}
\vspace{-3mm}
\end{table}

\subsection{Data Construction}

\subsubsection{Source Audio Data}
We selected source datasets according to two main criteria: audio quality and domain diversity. 
First, since MRMAD aims to evaluate fine-grained audio quality understanding, the original audio clips should be of sufficiently high quality so that the introduced degradations are perceptually distinguishable. 
Second, to ensure broad coverage, we collected audio clips from multiple datasets across three domains.
For speech, we sourced recordings of a diverse set of speakers from four high-quality speech corpora. 
Specifically, we included 108 speakers from VCTK~\cite{yamagishi2012english}, 80 from LibriSpeech \textit{dev-clean} and \textit{test-clean} subsets~\cite{panayotov2015librispeech}, 3 from HiFi-TTS \textit{clean} subset~\cite{bakhturina2021hi}, and 1 from LJSpeech~\cite{ljspeech17}. 
For each speaker, we sampled approximately 28 utterances, resulting in 5,400 speech clips in total. For music, we sampled songs from MUSDB18-HQ~\cite{MUSDB18HQ}, URMP~\cite{li2018creating}, and FMA-small~\cite{fma_dataset}. 
To improve musical diversity, we selected at most one song from each artist or composer.  
For each song, we randomly extracted two non-overlapping 15-second segments, yielding 4,800 music clips from 2,400 tracks.
For sound, we used TAU Urban Acoustic Scenes 2019~\citep{mesaros2019tau} and FSD50K~\citep{fonseca2021fsd50k}.
We sampled 5 clips from each of the 100 scene-location pairs in TAU 2019, and selected 3,700 high-quality, non-speech, non-music clips from a filtered FSD50K subset~\citep{yang2026fsd50k}.

Overall, we prepared 5,400 speech clips, 4,800 music clips, and 4,200 sound clips, following a 9:8:7 ratio that matches the number of degradation types in each domain.
This design balances clean-audio coverage with domain-specific degradation diversity.
For noise data, we collected background noise recordings from the DEMAND dataset~\citep{thiemann2013demand}.
All clips are converted to mono and resampled to 16~kHz to match the input format commonly used by current LALMs.

\subsubsection{Audio Degradations}
MRMAD is constructed by applying a diverse set of audio degradations to the prepared clips. For speech, we included all 9 degradation types.
For music, we excluded echo, since delayed repetitions can be intentional musical effects and are not always perceived as distortion. 
For sound, we excluded ambient noise and DSP denoising artifacts because many sound clips naturally contain ambience or noise-like content, making these degradations perceptually ambiguous.

Before applying degradations, each clip is first RMS-normalized to a consistent loudness level and saved as the reference audio. 
For each reference, we generated the degraded clips at three severity levels---\textit{audible}, \textit{medium}, and \textit{strong}---using different parameter settings. 
To keep severity perceptually comparable across domains, we slightly adjusted parameter ranges for speech, music, and sound, which differ in spectral structure, temporal dynamics, and perceptual sensitivity. Further details are provided in Appendix~\ref{app:degradation_details}.

\subsubsection{Question \& Prompt Generation}
We generated multiple-choice questions using a template-based prompting procedure.
Each question follows a multi-round format, where models receive one audio clip and one text prompt in each round.
To reduce prompt-specific bias, we prepared multiple natural-language templates and randomly sampled one per question.
We also appended an explicit answer constraint, requiring the model to output only a letter, such as A, B, C, or D. Examples and details are illustrated in Appendix~\ref{sec:template}.

\xhdr{Degradation Type Identification} DTI evaluates whether LALMs can identify a degradation type by comparing a degraded clip with its clean reference. 
Each DTI question has two turns: the first provides the reference audio, and the second provides the degraded audio with four candidate degradation types.
We use only the \textit{strong} level to ensure the artifact is clearly perceptible. Three distractors are sampled from valid degradation types outside the correct answer's perceptual group.
We categorize \{\textit{low-pass filtering, band-pass filtering, MP3 compression}\} as spectral or spectro-temporal distortions; \{\textit{reverberation, echo}\} as temporal effects; \{\textit{neural vocoder, DSP denoising}\} as algorithm-induced processing artifacts; ambient noise and packet loss as two separate groups.
This avoids overly ambiguous choices and encourages identification of the dominant degradation category. 

\xhdr{Degradation Severity Comparison}\ DSC evaluates whether LALMs can compare the severity of the same degradation across two clips. 
Each question consists of two turns: the first provides a degraded clip and its degradation type, and the second provides another clip with the same degradation type but a different severity level for comparison.
Both clips are generated from the same reference, with one at the \textit{audible} level and the other at the \textit{strong} level, ensuring a clear perceptual contrast.
For each instance, we randomized the order of the two clips and whether the prompt asks for higher or lower severity, reducing fixed response patterns. 

\xhdr{Degradation Severity Ranking}
DSR evaluates whether LALMs can rank three audio clips by the severity of the same degradation type. 
Each question has three turns, presenting three degraded versions of the same reference at \textit{audible}, \textit{medium}, and \textit{strong} levels.
The clips are randomly ordered across turns, and the prompt asks for either ascending or descending severity ranking.
Inspired by MuChoMusic~\citep{weck2024muchomusic}, the answer space contains four ranking options: the correct ranking, two close distractors formed by swapping adjacent positions, and one easier distractor sampled from the remaining incorrect permutations. 
This design evaluates fine-grained severity reasoning beyond identifying only the most or least degraded clip.

\subsubsection{Question \& Prompt Sampling}
After question generation, we sampled a balanced subset to ensure broad and controlled coverage in the final benchmark.
It contains 8,400 questions following a 3:3:1 split across DTI, DSC, and DSR, with each instance corresponding to a unique reference clip.
For speech, each task covers all 192 speakers with balanced questions per speaker.
We also assigned an equal number of questions to each degradation type and restricted each speaker-degradation pair to appear at most once per task.
For music, sampling is performed at the track level.
DTI and DSC are track-disjoint and jointly cover all 2,400 tracks.
To avoid segment overlap, DTI and DSC use one segment from each selected track, while DSR samples only a subset of the remaining segments.
Each task also maintains balanced degradation coverage.
For sound, sampling is performed at the clip level.
Each selected question uses a unique reference clip.
Similarly, we assigned an equal number of questions to each degradation for each task. Table~\ref{tab:dataset_summary} summarizes the overall statistics of the MRMAD benchmark.

\begin{table}[htbp]
\centering
\small
\setlength{\tabcolsep}{0.4pt}
\caption{Summary statistics of MRMAD, including question counts, degradation coverage, domain and task distributions, and average audio and question lengths.}
\label{tab:dataset_summary}
\begin{tabular}{lc}
\toprule
\textbf{Statistics}
& \textbf{Number} \\
\midrule
Total Questions & 8,400 \\
Audio domains & 3 \\
Degrad. Types (Speech : Music : Sound)  & \ 9 : 8 : 7 \\
\# Questions (Speech : Music : Sound) & \ 9 : 8 : 7 \\
\# Questions (DTI : DSC : DSR) & \ 3 : 3 : 1 \\
\midrule
Average Audio Length & 9.1 s \\
\# Speech : Music : Sound & 4.7 : 15.0 : 8.1 \\
\midrule
Average Question Length & 73.8 words \\
\# Speech : Music : Sound & 70.9 : 63.5 : 113.4 \\
\bottomrule
\end{tabular}
\end{table}

\section{Experimental Setup}
\label{sec:exps}

\begin{table*}[t]
\centering
\caption{\small Performance comparison of open- and closed-source audio-language models on MRMAD across three tasks and audio domains. Average scores are computed as correct predictions divided by total samples. Best results are highlighted in {\setlength{\fboxsep}{1.1pt}\colorbox{gray!20}{\textbf{\textit{bold}}}}.}
\vspace{-2mm}

\begingroup
\scriptsize
\setlength{\tabcolsep}{0.55pt}
\renewcommand{\arraystretch}{0.58}

\resizebox{0.965\textwidth}{!}{%
\begin{tabular}{@{}l c cccc cccc cccc@{}}
\toprule[1pt]
\midrule
\multirow{2}{*}{\textbf{Models}}
& \multirow{2}{*}{\textbf{Size}}
& \multicolumn{4}{c}{\textbf{DTI (Task 1, \%$\uparrow$)}}
& \multicolumn{4}{c}{\textbf{DSC (Task 2, \%$\uparrow$)}}
& \multicolumn{4}{c}{\textbf{DSR (Task 3, \%$\uparrow$)}} \\
\cmidrule(lr){3-6}
\cmidrule(lr){7-10}
\cmidrule(lr){11-14}
& & \textbf{Speech} & \textbf{Music} & \textbf{Sound} & \textbf{Avg.}
  & \textbf{Speech} & \textbf{Music} & \textbf{Sound} & \textbf{Avg.}
  & \textbf{Speech} & \textbf{Music} & \textbf{Sound} & \textbf{Avg.} \\
\midrule

\textcolor{gray}{Random Guess} & -- 
& \textcolor{gray}{25.00} & \textcolor{gray}{25.00} & \textcolor{gray}{25.00} & \textcolor{gray}{25.00} 
& \textcolor{gray}{50.00} & \textcolor{gray}{50.00} & \textcolor{gray}{50.00} & \textcolor{gray}{50.00} 
& \textcolor{gray}{25.00} & \textcolor{gray}{25.00} & \textcolor{gray}{25.00} & \textcolor{gray}{25.00}  \\

\midrule
\multicolumn{1}{l}{\underline{\textbf{Open-Source}}}
& \multicolumn{13}{c}{\textit{Large Audio Language Models (LALMs)}} \\
\midrule

SALMONN & 7B 
& 23.56 & 25.67 & 26.57 & 25.14 
& 49.11 & 50.58 & 49.14 & 49.61 
& 19.78 & 25.00 & 24.86 & 23.00 \\

Qwen2-Audio & 8.4B 
& 24.22 & 26.25 & 28.00 & 26.00 
& 41.78 & 45.42 & 50.48 & 45.53 
& 28.22 & 26.50 & 26.86 & 27.25 \\

Step-Audio-2-mini & 8B 
& 24.96 & 28.75 & 28.86 & 27.36 
& 52.96 & 50.08 & 50.67 & 51.33 
& 25.11 & 29.75 & 30.29 & 28.17 \\

Audio Flamingo 3 & 7B 
& 27.56 & 27.00 & 23.90 & 26.31 
& 51.19 & 48.83 &  {\setlength{\fboxsep}{1.1pt}\colorbox{gray!20}{\textbf{\textit{51.52}}}} & 50.50 
& 23.33 & 28.50 & 30.57 & 27.17 \\

Audio Flamingo Next & 8B 
& 26.00 & 28.00 & 23.52 & 25.94 
& 52.52 & 47.92 & 49.62 & 50.14 
& 25.78 & 31.00 & 31.71 & 29.25 \\

MiMo-Audio-Instruct & 7B
&  {\setlength{\fboxsep}{1.1pt}\colorbox{gray!20}{\textbf{\textit{36.22}}}} &  {\setlength{\fboxsep}{1.1pt}\colorbox{gray!20}{\textbf{\textit{32.83}}}} &  {\setlength{\fboxsep}{1.1pt}\colorbox{gray!20}{\textbf{\textit{31.05}}}} &  {\setlength{\fboxsep}{1.1pt}\colorbox{gray!20}{\textbf{\textit{33.58}}}} 
&  {\setlength{\fboxsep}{1.1pt}\colorbox{gray!20}{\textbf{\textit{55.11}}}} &  {\setlength{\fboxsep}{1.1pt}\colorbox{gray!20}{\textbf{\textit{51.25}}}} & 51.05 &  {\setlength{\fboxsep}{1.1pt}\colorbox{gray!20}{\textbf{\textit{52.64}}}}
&  {\setlength{\fboxsep}{1.1pt}\colorbox{gray!20}{\textbf{\textit{28.23}}}} &  {\setlength{\fboxsep}{1.1pt}\colorbox{gray!20}{\textbf{\textit{32.50}}}} &  {\setlength{\fboxsep}{1.1pt}\colorbox{gray!20}{\textbf{\textit{34.57}}}} &  {\setlength{\fboxsep}{1.1pt}\colorbox{gray!20}{\textbf{\textit{31.50}}}} \\

Voxtral-Mini & 4.7B 
& 24.07 & 27.33 & 25.71 & 25.64 
& 44.52 & 43.17 & 44.67 & 44.11 
& 25.78 & 27.75 & 31.71 & 28.17 \\

\midrule
\multicolumn{1}{l}{}
& \multicolumn{13}{c}{\textit{Large Audio Reasoning Models (LARMs)}} \\
\midrule

Step-Audio-R1 & 33B
& 29.78 & 28.17 & 30.48 & 29.44 
& 55.26 & 54.08 & 51.62 & 53.81 
& 27.11 & 28.50 & 27.43 & 27.67 \\

Audio-Reasoner & 8.4B 
& 18.81 & 23.08 & 24.57 & 21.92 
& 43.04 & 32.58 & 39.62 & 38.56 
& 15.78 & 10.75 & 14.29 & 13.67 \\

Qwen3-Omni-Thinking & 30B-A3B
&  {\setlength{\fboxsep}{1.1pt}\colorbox{gray!20}{\textbf{\textit{37.78}}}} &  {\setlength{\fboxsep}{1.1pt}\colorbox{gray!20}{\textbf{\textit{39.75}}}} &  {\setlength{\fboxsep}{1.1pt}\colorbox{gray!20}{\textbf{\textit{31.43}}}} &  {\setlength{\fboxsep}{1.1pt}\colorbox{gray!20}{\textbf{\textit{36.58}}}}
&  {\setlength{\fboxsep}{1.1pt}\colorbox{gray!20}{\textbf{\textit{61.48}}}} &  {\setlength{\fboxsep}{1.1pt}\colorbox{gray!20}{\textbf{\textit{61.50}}}} &  {\setlength{\fboxsep}{1.1pt}\colorbox{gray!20}{\textbf{\textit{58.75}}}} &  {\setlength{\fboxsep}{1.1pt}\colorbox{gray!20}{\textbf{\textit{60.70}}}} 
&  {\setlength{\fboxsep}{1.1pt}\colorbox{gray!20}{\textbf{\textit{27.56}}}} &  {\setlength{\fboxsep}{1.1pt}\colorbox{gray!20}{\textbf{\textit{34.50}}}} &  {\setlength{\fboxsep}{1.1pt}\colorbox{gray!20}{\textbf{\textit{36.29}}}} &  {\setlength{\fboxsep}{1.1pt}\colorbox{gray!20}{\textbf{\textit{32.42}}}} \\

Audio Flamingo 3-Think & 7B 
& 26.30 & 30.33 & 25.90 & 27.53 
& 51.26 & 49.83 & 51.52 & 50.86 
& 24.44 & 27.00 & 28.57 & 26.50 \\

Audio Flamingo Next-Think & 8B 
& 26.22 & 24.08 & 19.62 & 23.58 
& 49.63 & 48.42 & 52.48 & 50.06 
& 25.78 & 25.75 & 24.00 & 25.25 \\

MiMo-Audio-Think & 7B 
& 35.11 & 31.92 & 29.71 & 32.47 
& 46.81 & 49.00 & 48.48 & 48.03 
& 24.67 & 27.00 & 28.29 & 26.50 \\

\midrule
\multicolumn{1}{l}{}
& \multicolumn{13}{c}{\textit{Omni Language Models (OLMs)}} \\
\midrule

Qwen2.5-Omni & 10.7B 
& 27.33 & 30.50 & 29.90 & 29.14 
& 50.30 & 49.58 & 50.67 & 50.17 
&  {\setlength{\fboxsep}{1.1pt}\colorbox{gray!20}{\textbf{\textit{34.00}}}} &  {\setlength{\fboxsep}{1.1pt}\colorbox{gray!20}{\textbf{\textit{33.25}}}} &  {\setlength{\fboxsep}{1.1pt}\colorbox{gray!20}{\textbf{\textit{33.43}}}} &  {\setlength{\fboxsep}{1.1pt}\colorbox{gray!20}{\textbf{\textit{33.58}}}} \\

AudioMCQ-Weak-To-Strong & 10.7B 
& 27.19 & 28.33 & 27.90 & 27.78 
& 52.22 & 50.25 & 52.10 & 51.53 
& 31.78 & 30.25 & 32.86 & 31.58 \\

Omni-R1 & 10.7B 
& 27.85 & 29.75 & 30.29 & 29.19 
& 51.26 & 49.50 & 51.05 & 50.61 
& 28.22 & 29.25 & 31.14 & 29.42 \\

Qwen3-Omni-Instruct & 30B-A3B
&  {\setlength{\fboxsep}{1.1pt}\colorbox{gray!20}{\textbf{\textit{31.70}}}} &  {\setlength{\fboxsep}{1.1pt}\colorbox{gray!20}{\textbf{\textit{42.83}}}} &  {\setlength{\fboxsep}{1.1pt}\colorbox{gray!20}{\textbf{\textit{31.33}}}} &  {\setlength{\fboxsep}{1.1pt}\colorbox{gray!20}{\textbf{\textit{35.31}}}} 
&  {\setlength{\fboxsep}{1.1pt}\colorbox{gray!20}{\textbf{\textit{52.74}}}} &  {\setlength{\fboxsep}{1.1pt}\colorbox{gray!20}{\textbf{\textit{52.25}}}} &  {\setlength{\fboxsep}{1.1pt}\colorbox{gray!20}{\textbf{\textit{52.95}}}} &  {\setlength{\fboxsep}{1.1pt}\colorbox{gray!20}{\textbf{\textit{52.64}}}} 
& 22.00 & 27.25 & 29.43 & 25.92 \\

\midrule
\multicolumn{1}{l}{\underline{\textbf{Closed-Source}}}
& \multicolumn{13}{c}{\textit{LALMs \& LARMs \& OLMs}} \\
\midrule

GPT-Audio-1.5 & -- 
& 26.81 & 30.00 & 27.43 & 28.06
& 49.14 & 48.17 & 23.11 & 33.28 
& 23.11 & 24.75 & 20.29 & 22.83 \\

Gemini 3 Flash & --
& 59.63 & 43.92 & 41.05 & 48.97 
& 67.11 & 60.33 & 61.33 & 63.17 
& 40.67 & 39.50 & 38.57 & 39.67 \\

Gemini 3.1 Flash-Lite & -- 
& 36.89 & 32.92 & 30.57 & 33.72 
& 52.22 & 53.58 & 51.52 & 52.47 
& 34.00 & 30.25 & 31.14 & 31.92 \\

Gemini 3.1 Pro & -- 
&  {\setlength{\fboxsep}{1.1pt}\colorbox{gray!20}{\textbf{\textit{84.81}}}} &  {\setlength{\fboxsep}{1.1pt}\colorbox{gray!20}{\textbf{\textit{83.58}}}} &  {\setlength{\fboxsep}{1.1pt}\colorbox{gray!20}{\textbf{\textit{91.05}}}} &  {\setlength{\fboxsep}{1.1pt}\colorbox{gray!20}{\textbf{\textit{86.22}}}} 
&  {\setlength{\fboxsep}{1.1pt}\colorbox{gray!20}{\textbf{\textit{88.82}}}} &  {\setlength{\fboxsep}{1.1pt}\colorbox{gray!20}{\textbf{\textit{92.50}}}} &  {\setlength{\fboxsep}{1.1pt}\colorbox{gray!20}{\textbf{\textit{92.10}}}} &  {\setlength{\fboxsep}{1.1pt}\colorbox{gray!20}{\textbf{\textit{90.81}}}} 
&  {\setlength{\fboxsep}{1.1pt}\colorbox{gray!20}{\textbf{\textit{58.44}}}} &  {\setlength{\fboxsep}{1.1pt}\colorbox{gray!20}{\textbf{\textit{69.50}}}} &  {\setlength{\fboxsep}{1.1pt}\colorbox{gray!20}{\textbf{\textit{65.71}}}} & {\setlength{\fboxsep}{1.1pt}\colorbox{gray!20}{\textbf{\textit{64.25}}}} \\

Gemini 3.5 Flash & -- 
& 73.56 & 55.50 & 49.90 & 60.64 
& 84.30 & 76.50 & 72.00 & 78.11 
& 55.56 & 50.00 & 46.57 & 51.08 \\

\midrule
\bottomrule
\end{tabular}%
}

\endgroup

\label{tab:main_results}
\end{table*}

\begin{table*}[t]
\centering
\caption{\small Results of human evaluation and LALM performance. Average scores reflect the proportion of correct predictions relative to the total sample count. The best results are indicated in {\setlength{\fboxsep}{1.1pt}\colorbox{gray!20}{\textbf{\textit{bold}}}}, and the second-best ones in {\setlength{\fboxsep}{1.1pt}\colorbox{gray!20}{\textit{italics}}}.}

\begingroup
\scriptsize
\setlength{\tabcolsep}{0.55pt}
\renewcommand{\arraystretch}{0.58}

\resizebox{0.965\textwidth}{!}{%
\begin{tabular}{@{}l c cccc cccc cccc@{}}
\toprule[1pt]
\midrule
\multirow{2}{*}{\textbf{Models}}
& \multicolumn{4}{c}{\textbf{DTI (Task 1, \%$\uparrow$)}}
& \multicolumn{4}{c}{\textbf{DSC (Task 2, \%$\uparrow$)}}
& \multicolumn{4}{c}{\textbf{DSR (Task 3, \%$\uparrow$)}} \\
\cmidrule(lr){2-5}
\cmidrule(lr){6-9}
\cmidrule(lr){10-13}
& \textbf{Speech} & \textbf{Music} & \textbf{Sound} & \textbf{Avg.}
  & \textbf{Speech} & \textbf{Music} & \textbf{Sound} & \textbf{Avg.}
  & \textbf{Speech} & \textbf{Music} & \textbf{Sound} & \textbf{Avg.} \\
\midrule

\textcolor{gray}{Random Guess}
& \textcolor{gray}{25.00} & \textcolor{gray}{25.00} & \textcolor{gray}{25.00} & \textcolor{gray}{25.00} 
& \textcolor{gray}{50.00} & \textcolor{gray}{50.00} & \textcolor{gray}{50.00} & \textcolor{gray}{50.00} 
& \textcolor{gray}{25.00} & \textcolor{gray}{25.00} & \textcolor{gray}{25.00} & \textcolor{gray}{25.00}  \\

\midrule

Expert
& {\setlength{\fboxsep}{1.1pt}\colorbox{gray!20}{\textbf{\textit{100.00}}}} & {\setlength{\fboxsep}{1.1pt}\colorbox{gray!20}{\textbf{\textit{97.22}}}} & {\setlength{\fboxsep}{1.1pt}\colorbox{gray!20}{\textbf{\textit{98.41}}}} & {\setlength{\fboxsep}{1.1pt}\colorbox{gray!20}{\textbf{\textit{98.61}}}} 
& {\setlength{\fboxsep}{1.1pt}\colorbox{gray!20}{\textbf{\textit{100.00}}}} & {\setlength{\fboxsep}{1.1pt}\colorbox{gray!20}{\textbf{\textit{97.22}}}} & {\setlength{\fboxsep}{1.1pt}\colorbox{gray!20}{\textbf{\textit{100.00}}}} & {\setlength{\fboxsep}{1.1pt}\colorbox{gray!20}{\textbf{\textit{99.07}}}} 
& {\setlength{\fboxsep}{1.1pt}\colorbox{gray!20}{\textbf{\textit{100.00}}}} & {\setlength{\fboxsep}{1.1pt}\colorbox{gray!20}{\textbf{\textit{95.83}}}} & {\setlength{\fboxsep}{1.1pt}\colorbox{gray!20}{\textbf{\textit{100.00}}}} & {\setlength{\fboxsep}{1.1pt}\colorbox{gray!20}{\textbf{\textit{98.61}}}} \\

Non-Expert
& {\setlength{\fboxsep}{1.1pt}\colorbox{gray!20}{\textit{86.42}}} & 72.22 & 82.54 & 80.55 
& {\setlength{\fboxsep}{1.1pt}\colorbox{gray!20}{\textit{97.53}}} & {\setlength{\fboxsep}{1.1pt}\colorbox{gray!20}{\textit{93.06}}} & 90.48 & {\setlength{\fboxsep}{1.1pt}\colorbox{gray!20}{\textit{93.98}}} 
& {\setlength{\fboxsep}{1.1pt}\colorbox{gray!20}{\textit{88.89}}} & {\setlength{\fboxsep}{1.1pt}\colorbox{gray!20}{\textit{91.67}}} & {\setlength{\fboxsep}{1.1pt}\colorbox{gray!20}{\textit{80.95}}} & {\setlength{\fboxsep}{1.1pt}\colorbox{gray!20}{\textit{87.50}}} \\
\midrule
SALMONN
& 14.81 & 20.83 & 33.33 & 22.22 
& 40.74 & 54.17 & 52.38 & 48.61 
& 22.22 & 62.50 & 28.57 & 37.50 \\

Audio Flamingo 3 
& 25.93 & 25.00 & 19.05 & 23.61 
& 66.67 & 41.67 & 52.38 & 54.17 
& 22.22 & 75.00 & 28.57 & 41.67 \\

MiMo-Audio-Instruct
& 29.63 & 29.17 & 19.05 & 26.39 
& 44.44 & 45.83 & 42.86 & 44.44 
& 33.33 & 37.50 & 28.57 & 33.33 \\

Qwen3-Omni-Thinking
& 37.04 & 58.33 & 14.29 & 37.50 
& 66.67 & 75.00 & 52.38 & 65.28 
& 44.44 & 62.50 & 57.14 & 54.17 \\

Gemini 3.1 Pro 
& 85.19 & {\setlength{\fboxsep}{1.1pt}\colorbox{gray!20}{\textit{75.00}}} & {\setlength{\fboxsep}{1.1pt}\colorbox{gray!20}{\textit{90.48}}} & {\setlength{\fboxsep}{1.1pt}\colorbox{gray!20}{\textit{83.33}}} 
& 77.78 & 87.50 & {\setlength{\fboxsep}{1.1pt}\colorbox{gray!20}{\textit{100.00}}} & 87.50 
& 66.66 & 75.00 & 71.42 & 70.83 \\

\bottomrule
\end{tabular}%
}

\endgroup

\label{tab:human_results}
\end{table*}

\subsection{Models}
\label{sec:setups}
We evaluated 18 recent open-source and closed-source models on MRMAD, with each model following its official inference configuration.

\xhdr{LALMs}\ SALMONN~\citep{tang2023salmonn}, Qwen2-Audio~\citep{chu2024qwen2audio}, Step-Audio-2-mini~\citep{wu2025stepaudio2}, Audio Flamingo 3~\citep{goel2025audioflamingo3}, Audio Flamingo Next~\citep{ghosh2026audioflamingonext}, MiMo-Audio-Instruct~\citep{2025mimoaudio}, Voxtral-Mini~\citep{liu2025voxtral}, and GPT-Audio-1.5\footnote{\url{https://developers.openai.com/api/docs/models/gpt-audio-1.5}}.

\xhdr{LARMs}\ Step-Audio-R1~\citep{tian2025stepaudior1}, Audio-Reasoner~\citep{xie2025audioreasoner}, Qwen3-Omni-Thinking~\citep{xu2025qwen3omni}, Audio Flamingo 3-Think~\citep{goel2025audioflamingo3}, Audio Flamingo Next-Think~\citep{ghosh2026audioflamingonext} and MiMo-Audio-Think~\citep{2025mimoaudio}.

\xhdr{OLMs}\ Qwen2.5-Omni~\citep{xu2025qwen25omni}, AudioMCQ-Weak-To-Strong~\citep{he2025measuring}, Omni-R1~\citep{zhong2025omni}, Qwen3-Omni-Instruct~\citep{xu2025qwen3omni}, and various Gemini 3 models\footnote{\url{https://aistudio.google.com/models/gemini-3} and models are collected before 06/2026.}, including Gemini 3 Flash / 3.1 Flash-Lite / 3.1 Pro / 3.5 Flash~\citep{geminiteam2025}.

\subsection{Evaluation scheme}
We evaluate MRMAD with an output-based multiple-choice protocol.
For each instance, the model receives multi-round audio inputs, text prompts, and candidate options, and returns a single option label in the final turn.
All models use the same multi-turn inference templates and answer parser for fair comparison. Predictions are extracted by matching either the option label or a unique option string with rule-based regular expressions. Further details are provided in Appendix~\ref{sec:inference}.

\section{Results and Discussions}
\subsection{Main Results}
\label{sec:main_results}

Table~\ref{tab:main_results} presents the performance of open-source and closed-source models on MRMAD. Further analysis is provided in Appendix~\ref{sec:more_analysis}.
We summarize the main observations as follows.

\xhdr{MRMAD poses a substantial challenge to current LALMs}
Most open-source models perform only marginally above random guessing on MRMAD. 
On DTI and DSR, most models remain close to the 25\% random baseline, while on the DSC task, several models are near the 50\% binary-choice baseline. This indicates that current audio-language models, despite their progress on general audio understanding and reasoning, still struggle to capture low-level degradation cues.

\xhdr{Closed-source models lead, but the advantage is not uniform}
Gemini 3.1 Pro achieves the best performance across all three tasks, with 86.22\% on DTI, 90.81\% on DSC, and 64.25\% on DSR. 
Compared with the best open-source model on each task, it improves by 49.64, 30.11, and 30.67 absolute points, respectively. However, GPT-Audio-1.5 performs poorly, particularly on DSC, suggesting that closed-source deployment alone does not guarantee robust degradation-aware auditory perception.

\xhdr{Open-source models remain far from reliable audio degradation understanding}
Among open-source models, Qwen3-Omni-Thinking achieves the best average performance on DTI and DSC, with 36.58\% and 60.70\%, respectively. 
Qwen2.5-Omni achieves the strongest open-source DSR result, with 33.58\%. 
MiMo-Audio-Instruct is the best non-omni LALM, outperforming others across most tasks. 
Nevertheless, the large gap between these models and the Gemini 3 families indicates that existing open-source systems still lack reliable fine-grained perception and comparison capabilities of audio quality.

\begin{figure*}[t]
\centering
\includegraphics[width=0.75\textwidth]{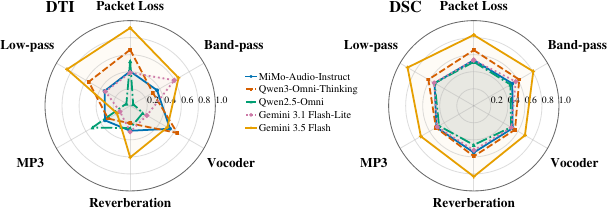}
\caption{Degradation-wise performance of five models on DTI and DSC, averaged over speech, music, and sound. 
DTI exhibits stronger artifact-specific variation, while DSC is more stable across degradation types.}
\label{fig:radar_dti_dsc}
\end{figure*}

\subsection{Reasoning vs. Non-Reasoning}
\label{sec:reasoning_vs_nonreasoning}

We further investigate whether explicit reasoning improves degradation-aware audio understanding by comparing matched reasoning and non-reasoning variants. 
As shown in Table~\ref{tab:main_results}, the effect of reasoning is mixed rather than consistently positive. 
Qwen3-Omni-Thinking improves over Qwen3-Omni-Instruct on all three tasks, with gains of 1.27 points on DTI, 8.06 points on DSC, and 6.50 points on DSR, suggesting that reasoning can benefit models when the underlying audio representation is sufficiently informative. 
However, this trend does not generalize across model families: Audio Flamingo3-Think yields only marginal differences, while Audio Flamingo Next-Think and MiMo-Audio-Think underperform their non-reasoning counterparts on most task averages. 
Averaged over matched pairs, reasoning impacts performance by only $-0.25$, $+0.93$, and $-0.79$ points on DTI, DSC, and DSR, respectively. 
Moreover, Audio-Reasoner performs substantially below several non-reasoning baselines, indicating that reasoning-oriented training alone does not guarantee robustness on MRMAD. 
These results show that audio degradation understanding cannot be reduced to high-level reasoning alone. Models need to first encode fine-grained acoustic evidence, preserve these representations reliably across turns, and compare subtle perceptual cues across multiple audio inputs and dialogue contexts.
Without degradation-sensitive audio representations, additional reasoning steps may amplify language priors rather than improve auditory grounding.

\subsection{Human Evaluations}
\label{sec:human_vs_lalms}
Although the labels in our benchmark are generated through controlled degradation rather than subjective annotation, human performance provides a valuable reference for perceptual difficulty and human-level perceptual capability. We therefore conduct a small-scale human evaluation on a subset of 168 questions (2\% of MRMAD), selected to preserve the original distributions across domains and tasks. Six participants are invited: three audio-domain experts with professional experience in audio analysis or related fields, and three non-expert judges representing general listeners without specialized audio expertise. All participants independently evaluate the subset after a brief training session covering one example of each degradation type and audio type, along with a few example questions for each task. As shown in Table~\ref{tab:human_results}, experts achieve accuracies of 98.61\%, 99.07\%, and 98.61\% across the three tasks, respectively, while non-expert listeners achieve 80.55\%, 93.98\%, and 87.50\%. Except for Gemini 3.1 Pro, all models perform substantially below human evaluators, highlighting a significant gap between human auditory perception and the capabilities of current LALMs. These results indicate that, although the tasks are tractable for human listeners, current LALMs remain substantially limited in their ability to distinguish fine-grained acoustic degradations.

\subsection{Degradation Type Analysis}
\label{sec:degradation_type_analysis}

Figure~\ref{fig:radar_dti_dsc} reveals that the DTI performance of different models varies substantially across degradation types.  
Stronger models, such as Gemini 3.5 Flash and MiMo-Audio-Instruct, perform better on packet loss and low-pass filtering, but exhibit lower accuracy on MP3 compression, band-pass filtering, and reverberation.
Packet loss and low-pass filtering provide more salient perceptual cues through temporal discontinuities and broad high-frequency attenuation, suggesting that these models are more sensitive to coarse degradation cues than to subtle spectral, spatial, or artifact-specific distortions.
In contrast, DSC is more uniform across degradation types, likely because the degradation category is given and the task reduces to within-type severity comparison. 
However, the persistent gap between closed-source and open-source models indicates that severity comparison still requires degradation-sensitive acoustic representations. Overall, the degradation-wise results show that MRMAD captures both task-level difficulty and artifact-specific weaknesses in current LALMs.

\begin{figure}[htbp]
\centering
\includegraphics[width=0.482\textwidth]{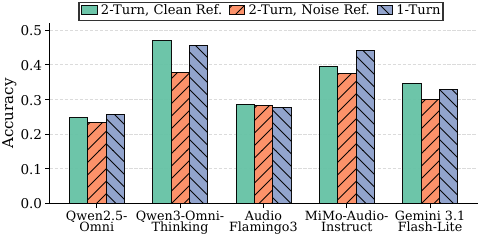}
\vspace{-3mm}
\caption{Performance on the DTI subset when the first-turn reference audio is either replaced with Gaussian noise or the entire first round is removed.}
\label{fig:ablation_accuracy}
\vspace{-2mm}
\end{figure}

\subsection{Why do LALMs Fail on MRMAD?}
\label{sec:failure_analysis}

\begin{figure*}[t]
\centering
\begin{subfigure}[t]{0.495\textwidth}
    \centering
    \includegraphics[width=\textwidth]{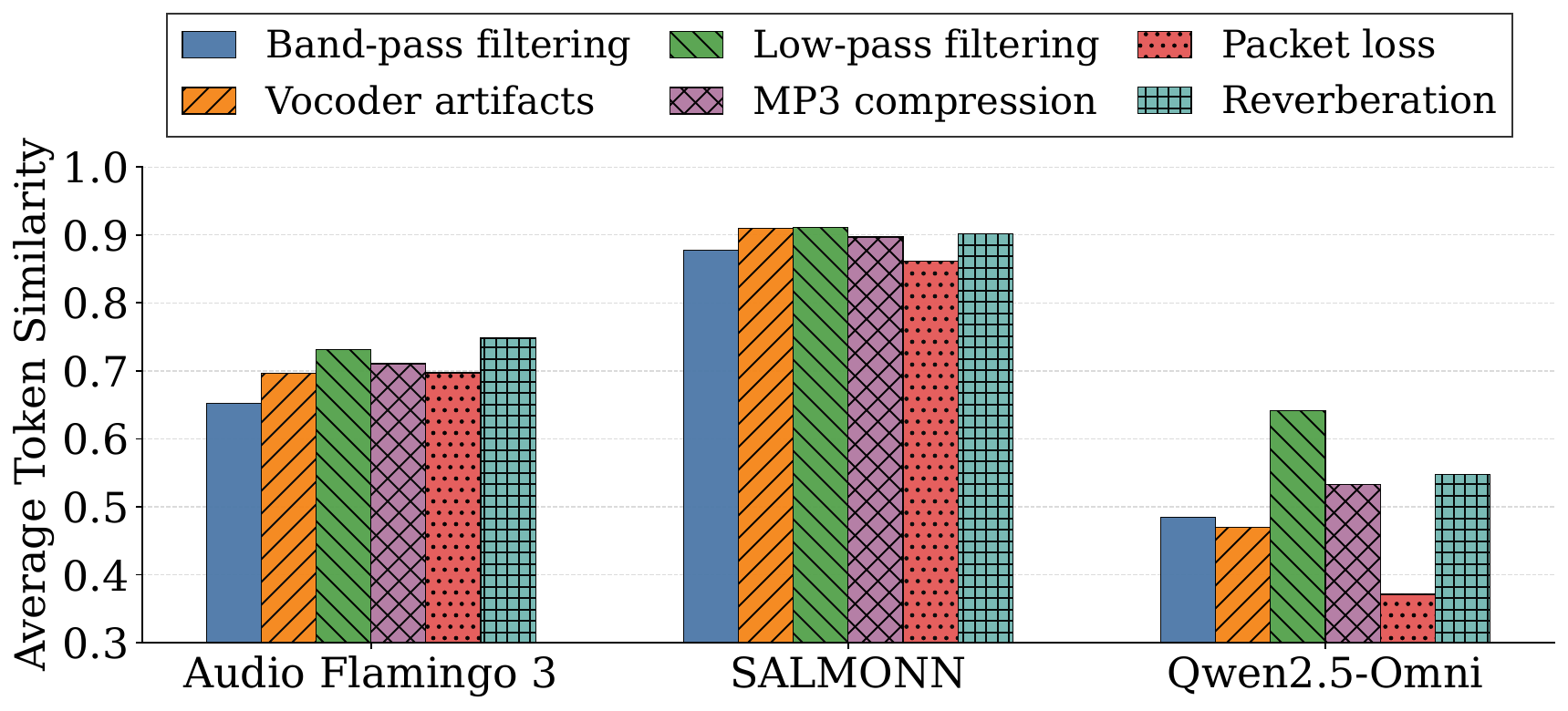}
    \caption{Average audio-token similarity between reference audio and strong-level degraded audio across six degradation types.}
    \label{fig:token_similarity_level3}
\end{subfigure}
\hfill
\begin{subfigure}[t]{0.495\textwidth}
    \centering
    \includegraphics[width=\textwidth]{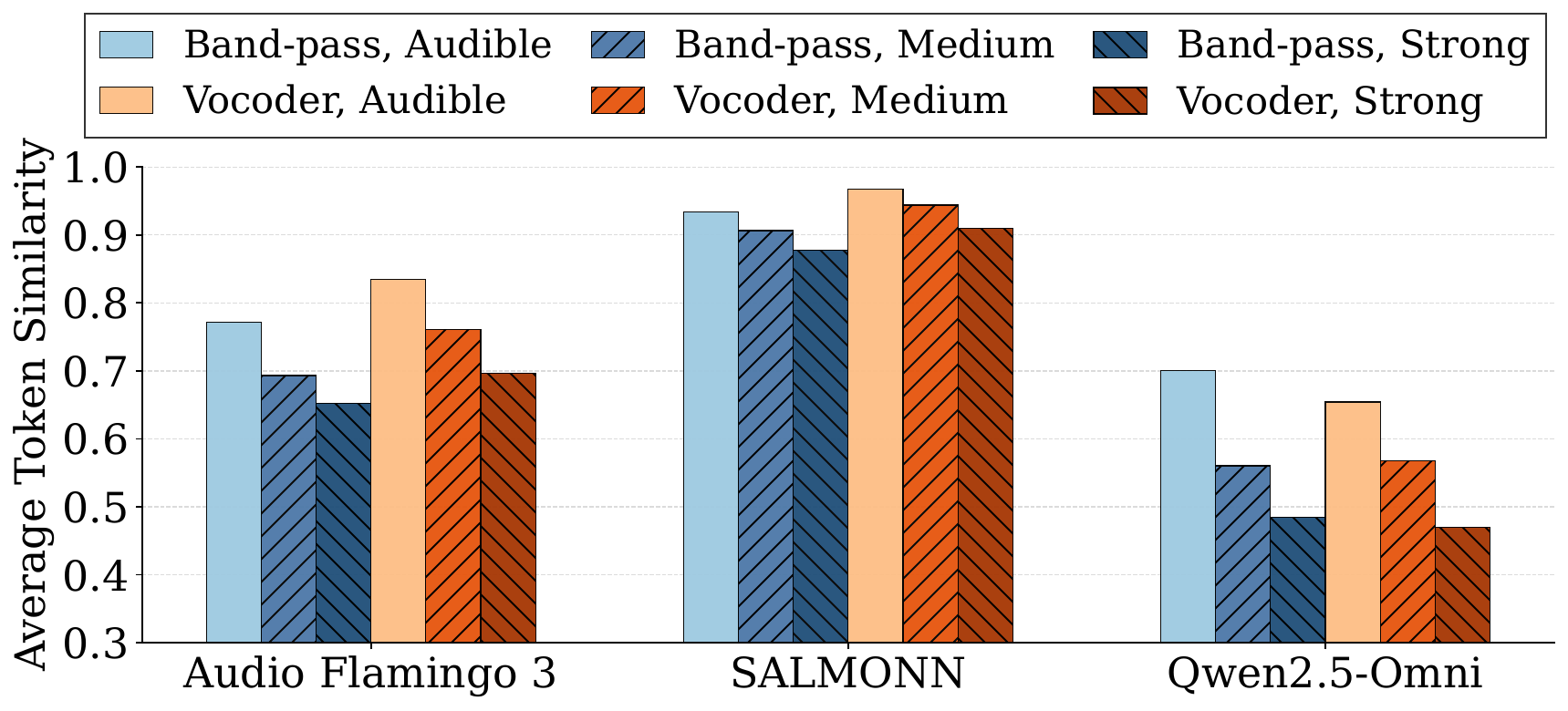}
    \caption{Average audio-token similarity for band-pass filtering and neural vocoder artifacts across three degradation severity levels.}
    \label{fig:token_similarity_lowpass}
\end{subfigure}
\caption{Token-level analysis of degradation-relevant information preservation in projected audio representations. Note that all three ALAMs apply Whisper-based audio encoders~\citep{radford2023robust}.}
\label{fig:token_similarity}
\vspace{-3.5mm}
\end{figure*}

We conduct controlled diagnostic analyses to identify potential failure modes in LALMs. 
We focus on DTI that compares the clean reference audio with its degraded counterpart. 
All diagnostics are conducted on a fixed and balanced subset of the DTI split, covering six shared degradation types across speech, music, and sound, with 900  samples. We select the representative models that reflect the overall performance trends.
We analyze three factors: whether models effectively use the reference audio, whether the multi-turn format introduces additional difficulty, and whether audio tokenizers preserve degradation-relevant information.

\xhdr{Clean reference audio is under-utilized}
In DTI, the clean reference in the first turn should provide essential evidence to identify the degradation in the second turn. 
We therefore compare the original setting with two variants: replacing the first-turn reference with Gaussian noise and even removing the first turn entirely. 
As shown in Figure~\ref{fig:ablation_accuracy}, corrupting or removing the reference causes only limited performance drops for most models. 
Qwen2.5-Omni and Audio Flamingo 3 remain nearly unchanged, while MiMo-Audio-Instruct even performs better in the one-turn setting when the clean reference is removed. 
Qwen3-Omni-Thinking shows the strongest dependence on the clean reference, but its one-turn result is still close to the original setting. 
These results suggest that many models do not reliably perform explicit reference-to-degraded comparison; instead, they often rely on the degraded clip alone or on language priors. 
Importantly, this study does not imply that models ignore audio entirely, but rather that they under-utilize the reference audio required by the task.

\xhdr{Multi-turn context is not consistently beneficial}
The one-turn variant removes the conversational dependency while preserving the degradation identification objective. 
If failures are mainly due to insufficient audio context, the original two-turn setting should consistently outperform the one-turn setting. 
However, Figure~\ref{fig:ablation_accuracy} shows that several models achieve comparable or higher accuracy when the first turn is deleted. 
This indicates that current LALMs may struggle to retain and compare audio evidence across turns, and that additional multi-turn context can introduce misalignment or memory burden rather than consistently improving reference-based reasoning~\citep{laban2025llms,xiao2026cantremember}.

\xhdr{Audio tokenizers often suppress degradation information} We further analyze whether the audio representations passed to the language model preserve sufficient degradation-relevant information. 
Given a reference clip $x$ and its degraded counterpart $\tilde{x}$, let 
$\mathbf{Z},\tilde{\mathbf{Z}}\in\mathbb{R}^{N\times d}$ denote their projected audio-token sequences after the final audio projection layer, where $N$ is the number of aligned tokens and $d$ is the embedding dimension. 
We compute the average token similarity as
$
\bar{s}(x,\tilde{x}) =
\frac{1}{N}\sum_{i=1}^{N}
\frac{\mathbf{z}_{i}^{\top}\tilde{\mathbf{z}}_{i}}
{\|\mathbf{z}_{i}\|_2\|\tilde{\mathbf{z}}_{i}\|_2}.
$
Lower similarity indicates greater sensitivity to degradation-induced acoustic changes, while consistently high similarity suggests invariance to low-level quality distortions. We consider models with audio encoders from the Whisper family~\citep{radford2023robust}.
As shown in Figure~\ref{fig:token_similarity}, SALMONN maintains high similarity between clean and strongly degraded audio across degradation types, suggesting that its projected tokens are largely invariant to artifact-level changes. 
Audio Flamingo 3 presents moderate sensitivity, whereas Qwen2.5-Omni exhibits lower and more degradation-dependent similarity. 
The degradation severity analysis is consistent with this pattern: Qwen2.5-Omni shows a clearer decrease in similarity score from audible to strong severity for band-pass filtering and vocoder artifacts, while SALMONN remains highly similar across severity levels. 
These results reveal that some LALMs may fail on MRMAD because degradation cues are attenuated before reaching the language model, limiting the evidence available for type identification and severity reasoning.

These diagnostics indicate that failures are from coupled grounding and representation bottlenecks. 
Models often under-utilize the clean reference and do not consistently benefit from multi-turn audio context, while their audio tokens may attenuate degradation cues before language decoding.

\section{Conclusion}
\label{sec:conclusion}

We propose MRMAD, a multi-round multi-audio benchmark for evaluating degradation-aware auditory understanding in LALMs across audio domains. 
It complements existing benchmarks by targeting low-level quality perception through DTI, DSC, and DSR. 
We show that current LALMs remain limited in fine-grained degradation perception and multi-audio severity reasoning, due to unstable reasoning gains, degradation-dependent behavior, weak cross-turn grounding, and degradation-insensitive audio representations. Furthermore, human evaluations reveal a clear limitation in their ability to identify acoustic degradation.
Therefore, MRMAD highlights degradation-aware perception as a critical missing capability for present audio-language modeling.

\section*{Limitations}
\label{sec:limitations}

First, MRMAD is designed as an evaluation benchmark, and we do not propose novel training methods or architectural solutions to improve LALMs. 
Accordingly, while our analysis identifies limitations in degradation-aware perception, cross-turn grounding, and audio-token representations, addressing these issues is left to future work. 
Second, several strong closed-source models, particularly the Gemini 3 family, are also among the strongest general-purpose multimodal models. 
Since their detailed architectures, training data, and audio-processing pipelines are not publicly available, we cannot fully disentangle whether their gains arise from audio-specific modeling, broader multimodal capabilities, scale, or proprietary training procedures. Then, the evaluated open-source LALMs do not invoke any external tools without agentic intervention~\cite{tong2026autagent,wang2026audio,wang2026hear,chen2026audiorouter}. Finally, due to the limited availability of expert human annotators for audio quality assessment, we only report human performance from a relatively small-scale evaluation to primarily quantify the discrepancy between LALMs and human auditory perception.

\section*{Ethics Statement}
\label{sec:ethics}

MRMAD is merely constructed from publicly available audio datasets and simulated degradations. 
The benchmark is intended for evaluating audio-language models on perceptual degradation understanding, not for identifying speakers, inferring private attributes, or making high-stakes decisions. 
All audio sources are used in accordance with their original licenses and usage terms.  
The benchmark may reveal weaknesses in current LALMs, but it does not introduce methods for generating harmful content or manipulating real-world audio systems.

\bibliography{refs}

\appendix
\appendix

\section*{Appendix}

\section{Degradation Simulation Details}
\label{app:degradation_details}
This section provides a detailed introduction to the nine degradation types used in MRMAD and their parameter configurations. 

\subsection{Ambient Noise}
\label{app:ambient_noise}

Ambient noise is simulated by mixing the reference audio with background noise recordings randomly sampled from the DEMAND dataset~\citep{thiemann2013demand}. 
For speech degradation, we exclude the \texttt{OMEETING} scene because it contains speech-like meeting recordings, which could introduce confounding speech content rather than purely background noise. 
The degradation severity is controlled by the signal-to-noise ratio (SNR), where lower SNR values indicate stronger noise corruption. 
Specifically, we sample SNRs from [10, 15]~dB for the audible level, [0, 5]~dB for the medium level, and [-10, -5]~dB for the strong level.

\subsection{Low-Pass Filtering}
\label{app:lowpass}
Low-pass filtering is simulated by attenuating high-frequency content above a sampled cutoff frequency. 
Lower cutoff frequencies correspond to stronger bandwidth limitation and thus more severe degradation. 
For speech, we sample cutoff frequencies from [4250, 5500]~Hz for the audible level, [2250, 3000]~Hz for the medium level, and [750, 1500]~Hz for the strong level. 
We use domain-specific cutoff ranges for music and sound because speech, music, and sound events differ in spectral bandwidth and perceptual reliance on high-frequency content. 
This keeps the three severity levels perceptually comparable across domains.

\subsection{Band-Pass Filtering}
\label{app:bandpass}
Band-pass filtering is simulated by preserving a frequency band around a sampled center frequency while attenuating frequencies outside the passband. 
Let \(f_c\) denote the center frequency and \(r_b\) denote the bandwidth fraction. 
The resulting passband is defined as
\begin{equation}
\left[f_c\left(1-\frac{r_b}{2}\right),\; f_c\left(1+\frac{r_b}{2}\right)\right],
\end{equation}
where smaller \(r_b\) values retain a narrower frequency range and therefore produce stronger degradation.

For each reference audio clip, we keep \(f_c\) fixed across the three severity levels and vary only \(r_b\). 
This isolates severity changes to bandwidth restriction, avoiding confounds from different affected frequency regions that may be perceived differently.
For speech and music, we sample \(r_b\) from [1.50, 1.75] for the audible level, [0.85, 1.25] for the medium level, and [0.30, 0.60] for the strong level. 
For sound events, we use larger \(r_b\)  at each severity level than those used for speech and music. 
This preserves more spectral content, since sound events often have diverse and sparse spectral patterns, making overly narrow band-pass filtering less consistently interpretable.

\subsection{MP3 Compression}
\label{app:mp3}

MP3 compression is simulated by encoding and decoding the reference audio using the \texttt{libmp3lame} codec. 
The degradation severity is controlled by the bitrate, where lower bitrates introduce stronger codec artifacts such as distortion, high-frequency loss, and temporal smearing. 
For all audio types, we use 32~kbps for the audible level, 16~kbps for the medium level, and 8~kbps for the strong level.

\subsection{Reverberation}
\label{app:reverberation}

Reverberation is simulated using algorithmic room effects with three parameters: room size, damping, and wet ratio. 
The room size controls the perceived spatial extent of the reverberant environment, damping controls the decay of high-frequency reflections, and the wet ratio controls the amount of reverberated signal mixed with the reference audio. 
To define the three severity levels, we increase the room size and wet ratio while decreasing damping, producing progressively stronger room coloration and longer reverberant tails.
We use the same parameter configuration for speech and music. 
For sound events, we use larger room size and wet ratio values, together with lower damping values, than those used for speech and music at the same level. 
This makes reverberation more perceptible for short or transient sound events, where weaker room effects may be less distinguishable.

\subsection{Echo}
\label{app:echo}

Echo is simulated using a multi-tap delay model. 
For each reference audio clip, we first sample a base delay, an initial tap weight, a decay factor, and the number of delay taps. 
The delay times are then set as integer multiples of the base delay, and the tap weights decay exponentially across taps. 
We keep these delay-related parameters fixed across the three severity levels and vary only the wet ratio, so that differences across levels are mainly caused by the amount of delayed signal rather than changes in echo timing.
The wet ratio is sampled from [0.20, 0.30] for the audible level, [0.50, 0.60] for the medium level, and [0.80, 0.90] for the strong level. 
Higher wet ratios introduce more prominent delayed repetitions, resulting in more severe echo degradation.

\subsection{Neural Vocoder Artifacts}
\label{app:hifigan}

Neural vocoder artifacts are simulated by reconstructing audio from corrupted log-mel representations. 
Given a reference waveform, we first convert it to a log-mel spectrogram, add Gaussian noise to the mel features, and then synthesize the waveform using a neural vocoder. 
The degradation severity is controlled by the standard deviation of the added mel-domain noise, denoted by \(\sigma_{\mathrm{mel}}\). 
Larger \(\sigma_{\mathrm{mel}}\) values introduce stronger distortions in the reconstructed waveform, producing artifacts such as spectral smearing, unnatural timbre, and vocoder-like reconstruction errors.

For speech, we employ a 16~kHz HiFi-GAN vocoder from SpeechBrain~\cite{speechbrain,speechbrain_v1}, which directly reconstructs waveforms from the mel-spectrograms. 
For music and sound events, we use a 24~kHz BigVGAN vocoder~\cite{lee2022bigvgan} because it is better suited for general audio reconstruction with broader spectral content. 
In this pipeline, the 16~kHz reference audio is first resampled to 24~kHz, converted to the mel-spectrograms, corrupted in the mel domain, reconstructed by BigVGAN at 24~kHz, and finally resampled back to 16~kHz. 
We use different \(\sigma_{\mathrm{mel}}\) ranges across audio types because the vocoders and source domains differ in spectral complexity and reconstruction robustness.

\subsection{DSP Denoising Artifacts}
\label{app:dsp_denoising}

DSP denoising artifacts are simulated by first adding background noise to the reference audio and then applying spectral-gating-based noise reduction~\cite{prasetio2024spectral}. 
Spectral gating estimates a frequency-dependent noise threshold from the spectrogram and applies a mask to suppress components below the threshold. 
While this reduces background noise, aggressive suppression can introduce perceptual artifacts such as over-suppression, signal distortion, and musical noise.

The degradation severity is mainly controlled by the noise-reduction strength \(\alpha\). 
Larger \(\alpha\) values suppress more estimated noise but also introduce stronger denoising artifacts. 
We sample \(\alpha\) from [0.950, 0.960] for the audible level, [0.980, 0.985] for the medium level, and [0.997, 1.000] for the strong level. 
For each reference clip, we use the same added noise segment and the same SNR across the three severity levels, so that severity differences are primarily caused by the denoising strength rather than differences in the noise condition. 
Other denoising parameters, including the time constant and frequency-mask smoothing, are also shared across severity levels.

\subsection{Packet Loss}
\label{app:packet_loss}

Packet loss is simulated by dividing the reference audio into fixed-length packets and dropping selected packets according to a two-state Markov process~\cite{lin2021time}. 
Each packet is assigned to either a normal state \(N\) or a loss state \(L\). 
The parameter \(p_N\) denotes the probability of remaining in the normal state, while \(p_L\) denotes the probability of remaining in the loss state. 
The theoretical packet loss rate is given as
\begin{equation}
\rho = \frac{1-p_N}{(1-p_N)+(1-p_L)}.
\end{equation}
When a packet is marked as lost, its samples are replaced with zeros, producing discontinuities similar to missing audio frames in packet-based transmission. 
We use 30~ms packets for all audio types. 
To construct the three severity levels, we choose lower \(p_N\) values and higher \(p_L\) values for stronger degradations, which increases both the frequency of loss events and the duration of loss bursts.

\section{Additional Analysis}
\label{sec:more_analysis}
\xhdr{Severity ranking is the most difficult task}
DSR yields the lowest scores for nearly all models. Even the best open-source model reaches only 33.58\%, slightly above the 25\% random baseline, and the best closed-source model drops substantially from its DTI and DSC performance. 
This suggests that ranking multiple degraded clips requires more than detecting the presence of an artifact; models must preserve severity-sensitive acoustic evidence across multiple audio inputs and conversational turns.

\xhdr{Model scaling law alone does not explain performance}
Although larger models such as Qwen3-Omni-Thinking and Step-Audio-R1 generally perform better than several smaller baselines, parameter count is not a sufficient predictor of MRMAD performance. 
MiMo-Audio-Instruct remains competitive with 7B parameters, while Audio-Reasoner performs substantially below several models of similar or smaller scale. 
This suggests that MRMAD evaluates capabilities that depend not only on model size but also on degradation-sensitive audio representations, training scale, and the ability to ground language outputs in acoustic evidence.

\section{Multi-turn Inference Template}
\label{sec:inference}
Figure~\ref{fig:inference_templates} shows the two-turn and three-turn templates for multi-round multi-audio evaluations. In the two-turn setting, the model first receives the Turn-1 audio $a_1$ together with its text prompt $t_1$ and generates an intermediate response $r_1$. The second turn preserves initial textual and audio inputs and the model's response as conversation history, then provides the Turn-2 audio $a_2$ and prompt $t_2$ for the final generation. The three-turn template extends this procedure by adding another audio-text pair $(a_3,t_3)$ while retaining all previous user and assistant turns. This design enables the model to accumulate information across multiple audio clips before making the last decision. The final response is parsed by $\mathrm{ExtractChoice}(\cdot)$ to obtain the predicted option $\hat{y}$ from the candidate set $\mathcal{Y}$.

\begin{figure*}[t]
\centering

\begin{minipage}[t]{0.48\textwidth}
\begin{algorithm}[H]
\caption{Two-turn inference template}
\label{alg:two_turn_template}
\begin{spacing}{1.1866}
\begin{algorithmic}[1]
\Require Turn-1 audio $a_1$, Turn-1 text prompt $t_1$
\Require Turn-2 audio $a_2$, Turn-2 text prompt $t_2$
\Require LALM $\mathcal{M}$, candidate option set $\mathcal{Y}$

\Statex
\State \textbf{Turn 1}
\State Construct Turn-1 conversation:
\[
\mathcal{C}_1 =
\big[
    \mathrm{User}(a_1, t_1)
\big].
\]
\State Generate Turn-1 response:
\[
r_1 = \mathcal{M}(\mathcal{C}_1).
\]

\Statex
\State \textbf{Turn 2}
\State Construct Turn-2 conversation by preserving Turn-1 context:
\[
\begin{aligned}
\mathcal{C}_2 =
\big[&
    \mathrm{User}(a_1, t_1),
    \mathrm{Assistant}(r_1),\\
&   \mathrm{User}(a_2, t_2)
\big].
\end{aligned}
\]
\State Generate Turn-2 response:
\[
r_2 = \mathcal{M}(\mathcal{C}_2).
\]

\Statex
\State Extract the predicted option:
\[
\hat{y} = \mathrm{ExtractChoice}(r_2), \quad \hat{y} \in \mathcal{Y}.
\]
\State \Return $\hat{y}, r_1, r_2$
\end{algorithmic}
\end{spacing}
\end{algorithm}
\end{minipage}
\hfill
\begin{minipage}[t]{0.48\textwidth}
\begin{algorithm}[H]
\caption{Three-turn inference template}
\label{alg:three_turn_template}
\begin{spacing}{0.75}
\begin{algorithmic}[1]
\Require Turn-1 audio $a_1$, Turn-1 text prompt $t_1$
\Require Turn-2 audio $a_2$, Turn-2 text prompt $t_2$
\Require Turn-3 audio $a_3$, Turn-3 text prompt $t_3$
\Require LALM $\mathcal{M}$, candidate option set $\mathcal{Y}$

\Statex
\State \textbf{Turn 1}
\State Construct Turn-1 conversation:
\[
\mathcal{C}_1 =
\big[
    \mathrm{User}(a_1, t_1)
\big].
\]
\State Generate Turn-1 response:
\[
r_1 = \mathcal{M}(\mathcal{C}_1).
\]

\Statex
\State \textbf{Turn 2}
\State Construct Turn-2 conversation by preserving Turn-1 context:
\[
\begin{aligned}
\mathcal{C}_2 =
\big[&
    \mathrm{User}(a_1, t_1),
    \mathrm{Assistant}(r_1),\\
&   \mathrm{User}(a_2, t_2)
\big].
\end{aligned}
\]
\State Generate Turn-2 response:
\[
r_2 = \mathcal{M}(\mathcal{C}_2).
\]

\Statex
\State \textbf{Turn 3}
\State Construct Turn-3 conversation by preserving previous contexts:
\[
\begin{aligned}
\mathcal{C}_3 =
\big[&
    \mathrm{User}(a_1, t_1),
    \mathrm{Assistant}(r_1),\\
&   \mathrm{User}(a_2, t_2),
    \mathrm{Assistant}(r_2),\\
&   \mathrm{User}(a_3, t_3)
\big].
\end{aligned}
\]
\State Generate Turn-3 response:
\[
r_3 = \mathcal{M}(\mathcal{C}_3).
\]

\Statex
\State Extract the predicted option:
\[
\hat{y} = \mathrm{ExtractChoice}(r_3), \quad \hat{y} \in \mathcal{Y}.
\]
\State \Return $\hat{y}, r_1, r_2, r_3$
\end{algorithmic}
\end{spacing}
\end{algorithm}
\end{minipage}

\caption{Two-turn and three-turn inference templates for multi-turn audio-language evaluation.}
\label{fig:inference_templates}
\end{figure*}

\begin{figure*}[htbp]
    \centering

    \begin{subfigure}[t]{0.495\textwidth}
        \centering
        \includegraphics[width=\linewidth]{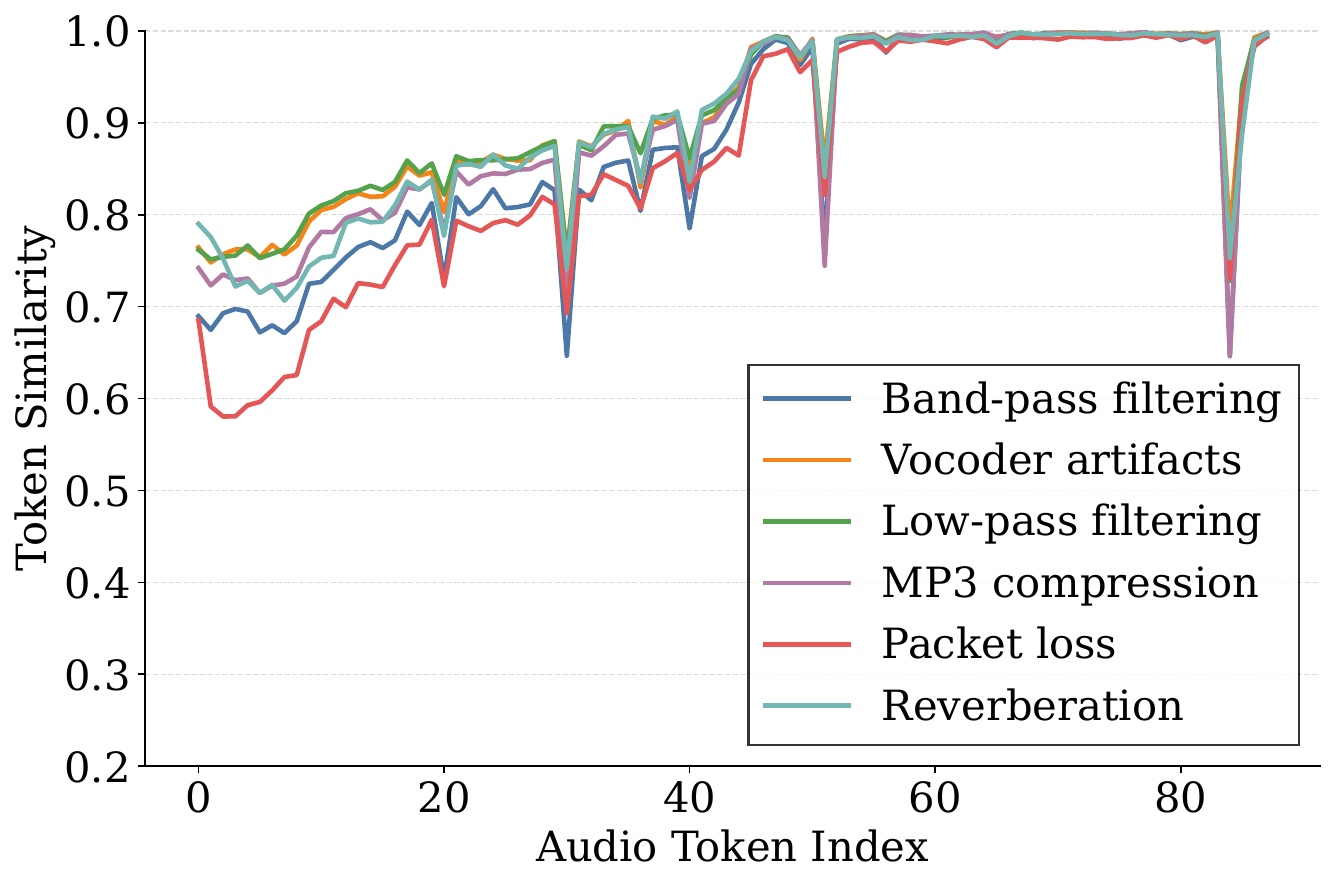}
        \caption{SALMONN.}
        \label{fig:left_subfig}
    \end{subfigure}
    \hfill
    \begin{subfigure}[t]{0.495\textwidth}
        \centering
        \includegraphics[width=\linewidth]{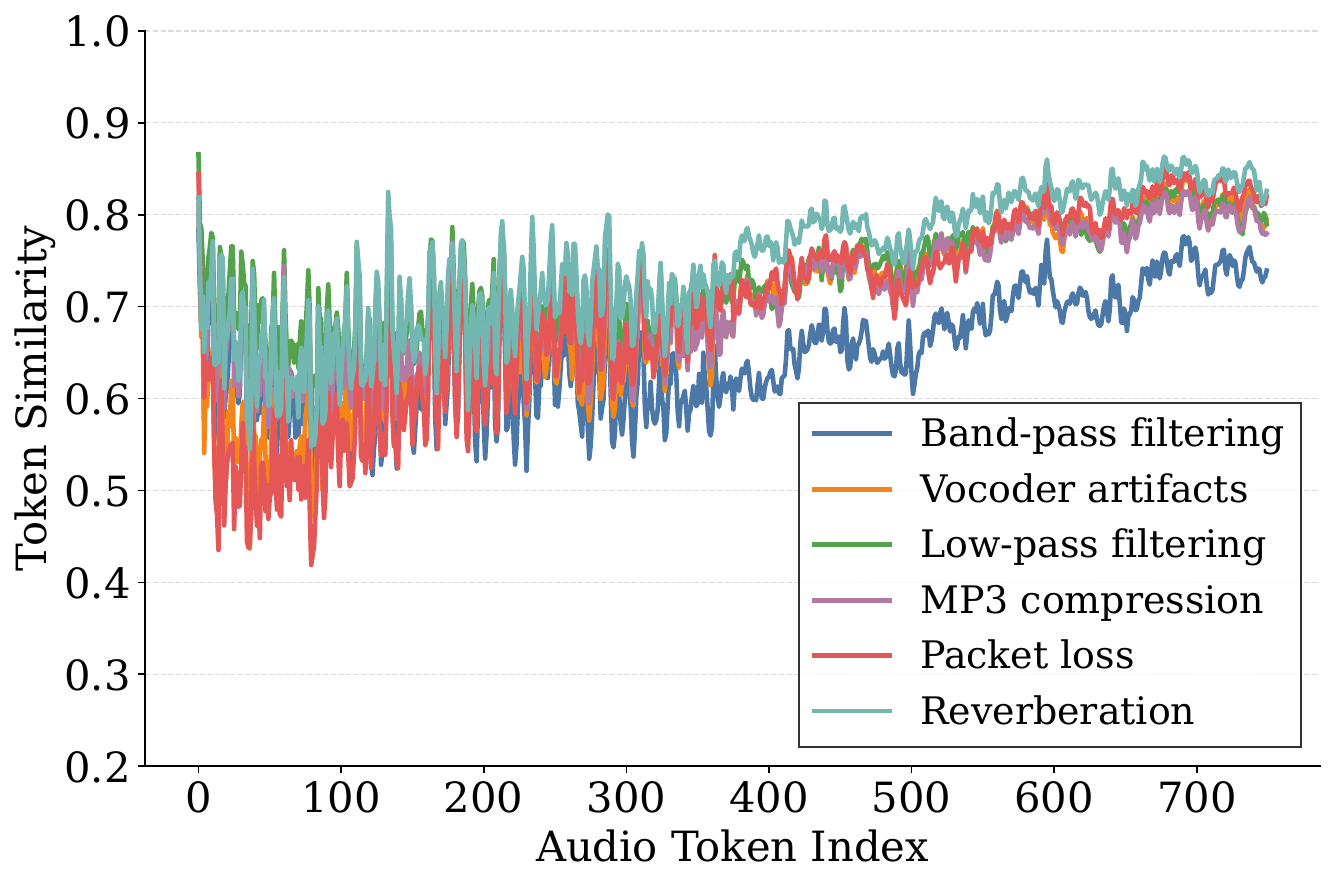}
        \caption{Audio Flamingo 3.}
        \label{fig:right_subfig}
    \end{subfigure}

    \caption{Projected audio-token similarity curves across six degradation types at the strong severity level.}
    \label{fig:strong_level_sim}
\end{figure*}

\begin{figure*}[htbp]
    \centering

    \begin{subfigure}[t]{0.495\textwidth}
        \centering
        \includegraphics[width=\linewidth]{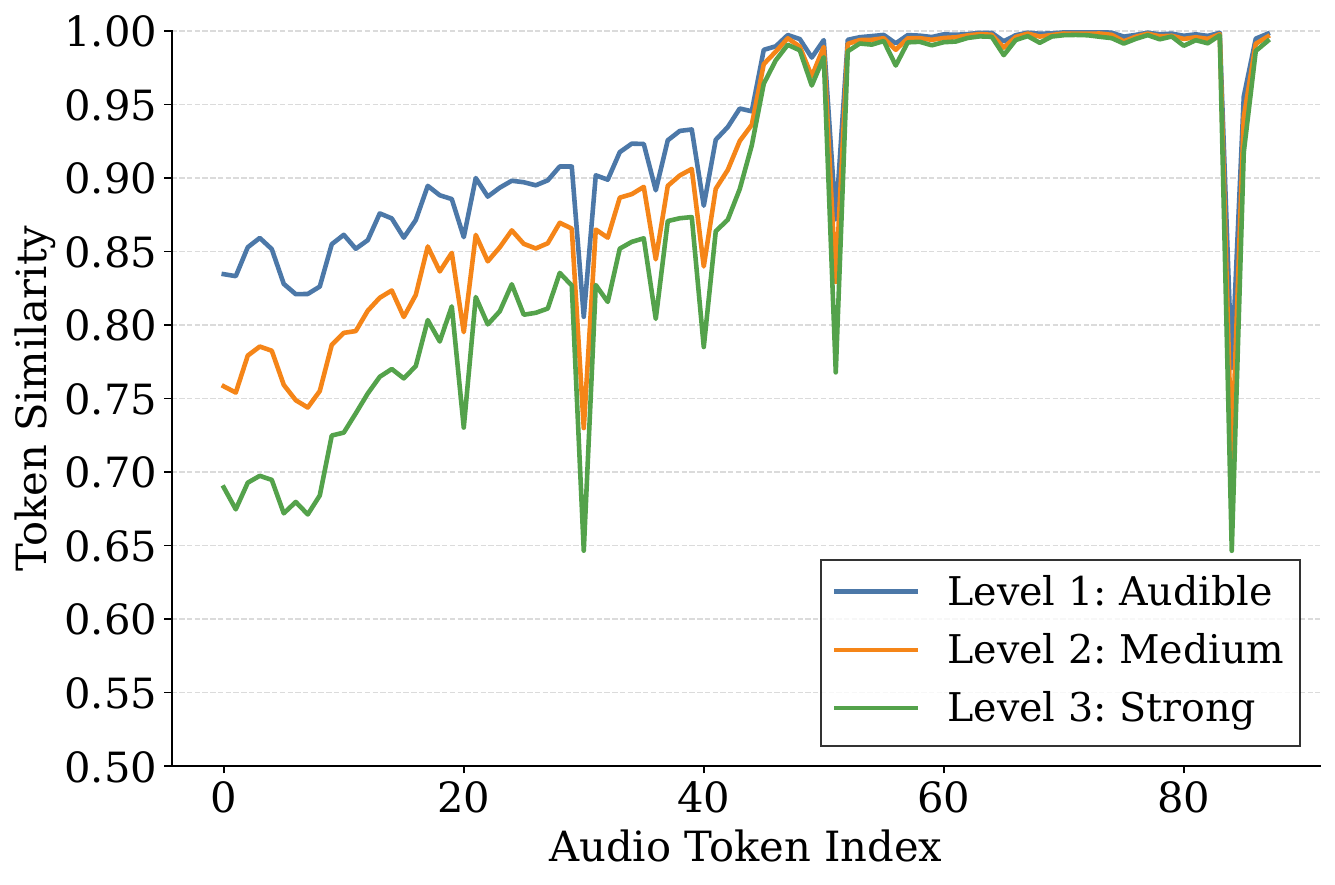}
        \caption{Band-pass filtering.}
        \label{fig:left_subfig}
    \end{subfigure}
    \hfill
    \begin{subfigure}[t]{0.495\textwidth}
        \centering
        \includegraphics[width=\linewidth]{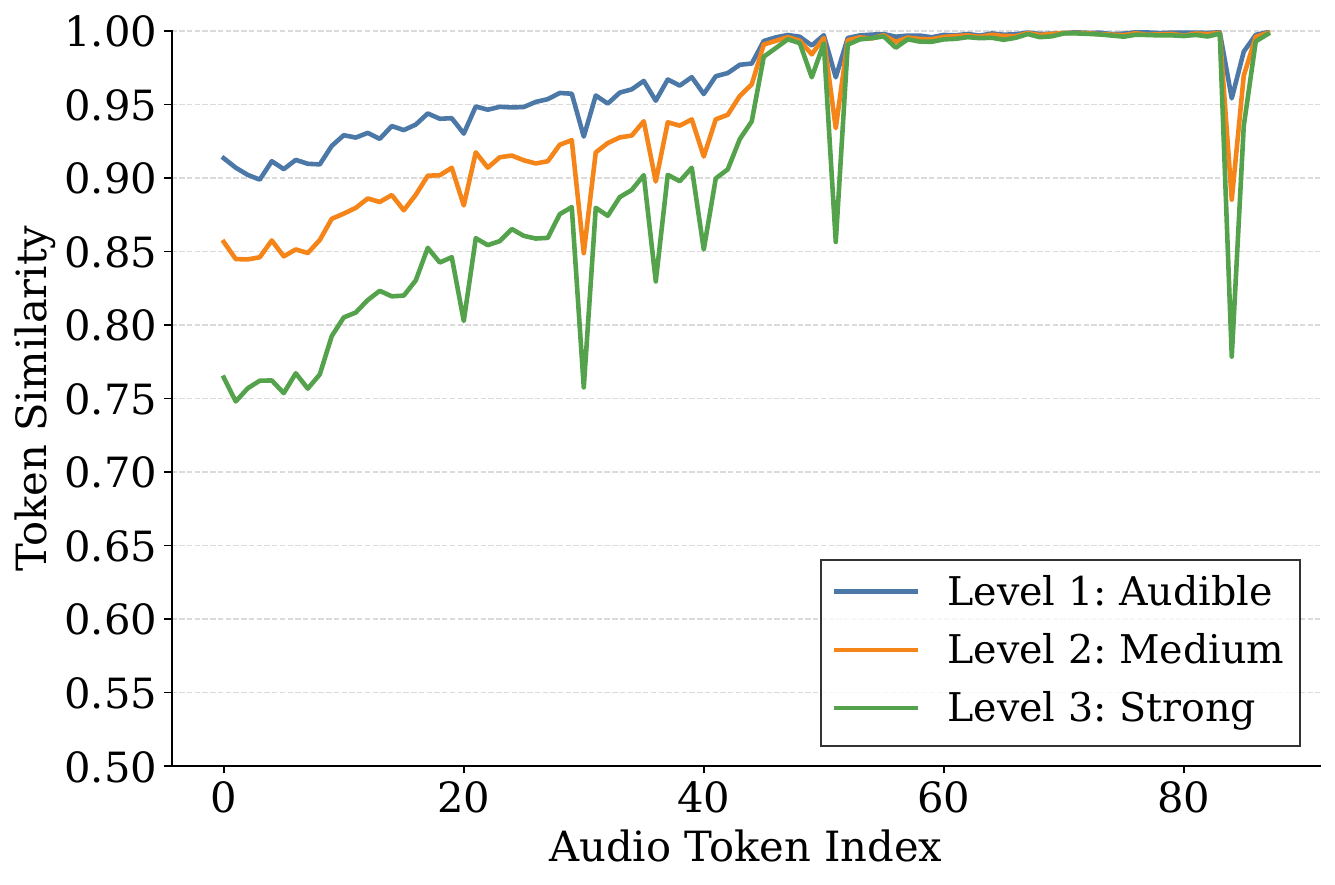}
        \caption{Neural vocoder artifacts.}
        \label{fig:right_subfig}
    \end{subfigure}

    \caption{Projected audio-token similarity curves across three severity levels for SALMONN.}
    \label{fig:SALMONN}
\end{figure*}

\begin{figure*}[htbp]
    \centering

    \begin{subfigure}[t]{0.495\textwidth}
        \centering
        \includegraphics[width=\linewidth]{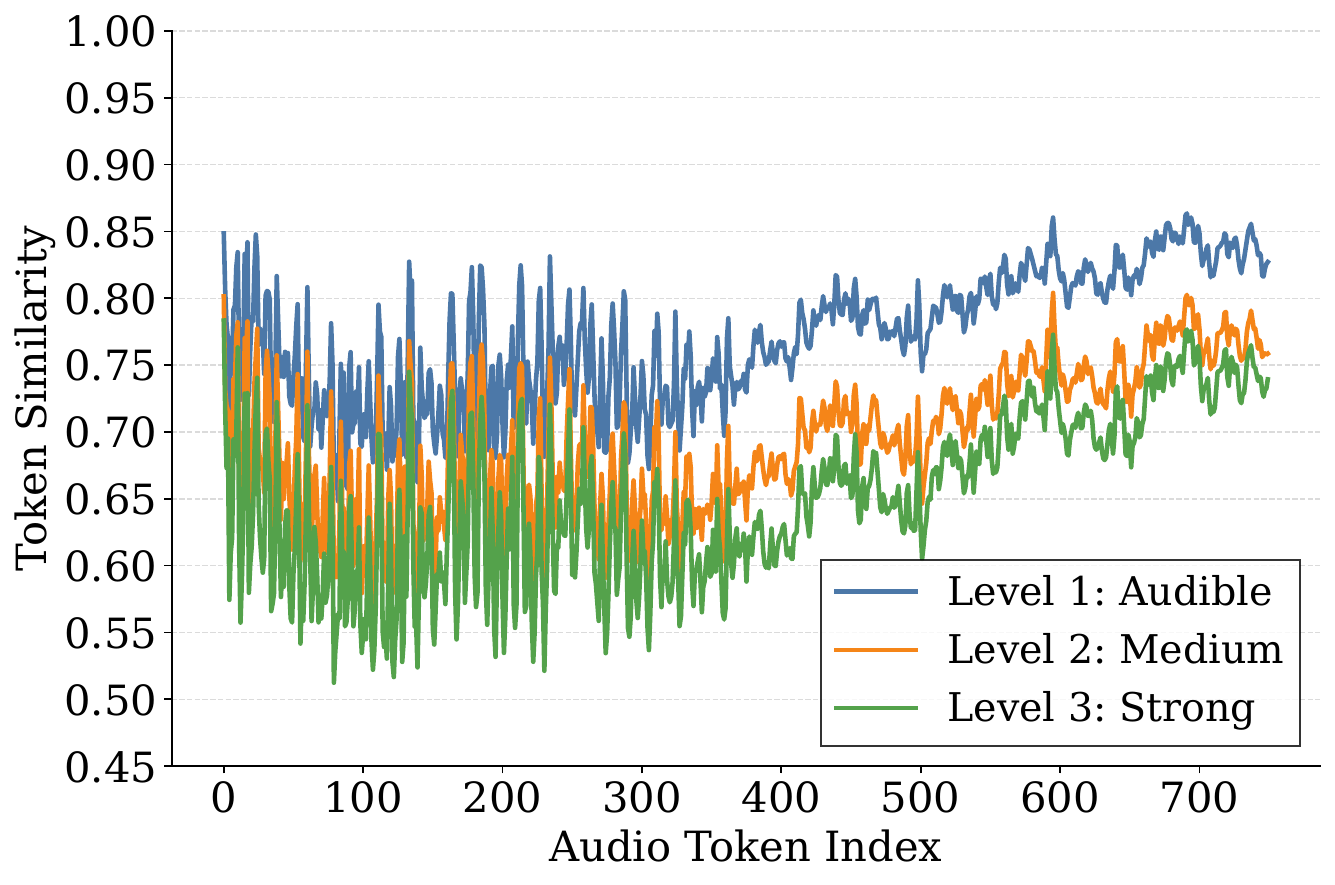}
        \caption{Band-pass filtering.}
        \label{fig:left_subfig}
    \end{subfigure}
    \hfill
    \begin{subfigure}[t]{0.495\textwidth}
        \centering
        \includegraphics[width=\linewidth]{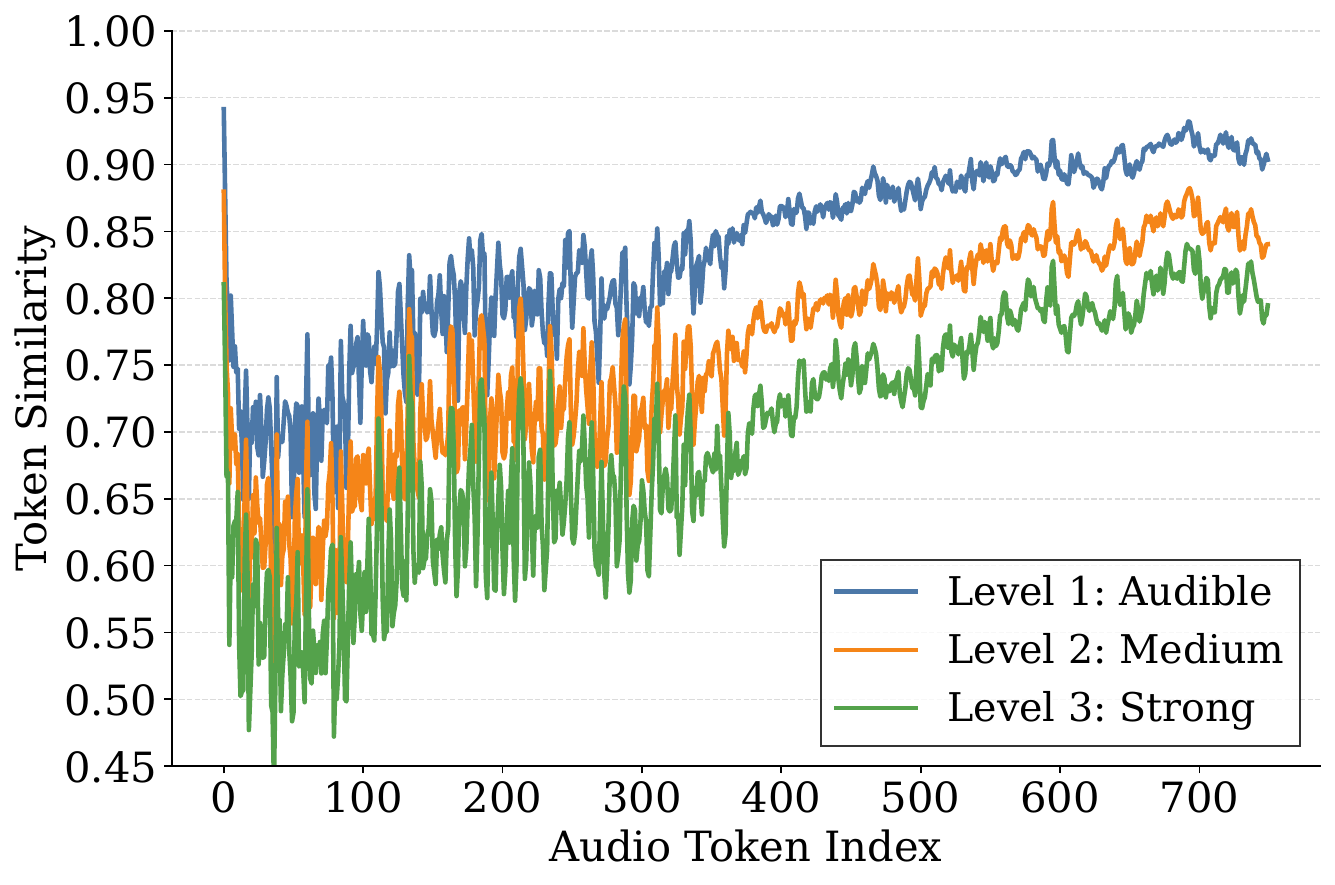}
        \caption{Neural vocoder artifacts.}
        \label{fig:right_subfig}
    \end{subfigure}

    \caption{Projected audio-token similarity curves across three severity levels for Audio Flamingo 3.}
    \label{fig:af3}
\end{figure*}

\section{Token-level Similarity}
\label{sec:token_similarity}
\xhdr{Similarity across degradation types}
Figure~\ref{fig:strong_level_sim} shows token-level similarity curves between clean and strongly degraded audio across six degradation types. SALMONN exhibits consistently high similarity for most tokens, with curves for different degradations quickly converging to near-identical values. This suggests that its projected tokens largely preserve high-level audio content while attenuating degradation-specific differences. In contrast, Audio Flamingo 3 shows lower and more variable similarity across token positions and degradation types, indicating that its representation retains more degradation-induced variation. However, the curves remain noisy and partially overlapping, suggesting that degradation categories are not clearly separated at the token level.

\xhdr{Similarity across degradation severity levels}
Figures~\ref{fig:SALMONN} and~\ref{fig:af3} compare token similarity across audible, medium, and strong severity levels for band-pass filtering and neural vocoder artifacts. For SALMONN, severity ordering appears only in the early tokens: stronger degradations yield lower similarity near the beginning of the sequence, but all severity levels quickly saturate toward high similarity. This suggests that severity information may be compressed or lost after projection, limiting the evidence available for downstream severity reasoning. Audio Flamingo 3 shows a clearer and more persistent monotonic trend, with stronger degradations generally corresponding to lower token similarity across much of the sequence. Nevertheless, the separation remains imperfect and varies across token positions, indicating that severity information is only partially and non-uniformly encoded.

Overall, the token-level analysis supports the failure patterns observed in MRMAD. 
Models with highly invariant projected tokens may lack sufficient low-level evidence for DTI and DSR, even when the text prompt explicitly asks for such judgments. 
Thus, improving degradation-aware LALMs requires audio encoders and projection modules that preserve fine-grained spectral, temporal, and degradation-specific cues rather than only semantic audio content.

\section{Response Examples}
\label{sec:examples}
Tables~\ref{tab:dti_response_examples}--\ref{tab:dsr_response_examples} show representative examples of correct and incorrect model responses across the DTI, DSC, and DSR tasks. The correct examples suggest that models succeed when their intermediate descriptions are aligned with the perceptual cues required by the task. For DTI, the model links the loss of high-frequency detail and muffled timbre to low-pass filtering. For DSC, it compares both clips along the same degradation dimension and recognizes that the second speech clip contains louder and more intrusive ambient noise. For DSR, it preserves severity information across three turns and correctly orders the clips according to the strength of the low-pass filtering artifact.

The incorrect examples reveal several limitations. In the DTI task, Qwen2-Audio loses task grounding and produces an irrelevant response, leading to an incorrect degradation-type prediction. In the DSC task, even the detailed reasoning produced by Qwen3-Omni-Thinking can disagree with the annotated severity, especially for complex neural-vocoder artifacts whose perceptual cues may be ambiguous. In the DSR task, Gemini 3.1 Pro can correctly describe individual clips but still sometimes fails to maintain a consistent global ranking or map the final ordering to the correct option. These results indicate that multi-round degradation reasoning requires not only artifact recognition, but also stable cross-turn memory, calibrated severity perception, and accurate option selection.

\section{Question \& Prompt Template}
\label{sec:template}
To reduce prompt-specific bias and repeated wording patterns, we adopt diversified templates for question generation. This design is motivated by prior work showing that LLM behavior is sensitive to adversarial prompt variations, prompt wording, structured query forms, and structured reasoning formulations~\citep{zhu2023promptrobust,xiong2024large,sclar2024quantifying,zou2025queryattack,liang2025embedding,hu2025loc,zou2026smartbench,xiong2026adaptive,xiong2026enhancing,li2026rethinking}. We therefore use task-specific prompt pools to encourage models to rely more on audio perception and task reasoning rather than superficial textual cues. The initial template pools are generated using ChatGPT-5.4 Thinking\footnote{\url{https://chat.openai.com/}} and then manually reviewed and curated, following recent work that leverages LLMs and multimodal LLMs for task-specific annotation, data synthesis, and evaluation construction.~\citep{long2024llms,hou2024improving,yang2025retrieval,wang2025filter,fu2025mallm,peng2026multifinben}.

The DTI templates are provided in Table~\ref{tab:dti_templates}, the DSC templates in Tables~\ref{tab:dsc_general_templates} and~\ref{tab:dsc_round2_templates}, and the DSR templates in Tables~\ref{tab:dsr_general_templates} and~\ref{tab:dsr_round3_templates}. For each instance, \texttt{\{audio\_type\}} is determined by the audio domain, while \texttt{\{degradation\_label\}} is determined by the audio's degradation type. For DTI, the final Round~1 text prompt string is directly sampled from the Round~1 text prompt pool, while the final Round~2 text prompt string is formed by concatenating a sampled Round~2 text prompt with a sampled answer constraint.
The choices \texttt{\{A\}}--\texttt{\{D\}} contain the correct degradation type and three distractors sampled from degradation types outside the correct answer's perceptual group, with option orders randomized to avoid answer-position bias.

For DSC, the final Round~1 text prompt string is sampled from the Round~1 prompt pool, with the degradation type explicitly mentioned to guide the comparison. The final Round~2 text string concatenates a sampled Round~2 introduction prompt, a degradation-specific prompt, the fixed option format, and a sampled answer constraint. The comparison phrase \texttt{\{cmp\}} is randomly sampled from the higher- or lower-severity phrase pool, and the order of the two clips is randomized to avoid positional and answer-option bias. Since different degradations involve distinct perceptual cues, such as high-frequency attenuation for low-pass filtering, reverberant tails for reverberation, and temporal discontinuities for packet loss, we use diversified degradation descriptions instead of repeatedly using only \texttt{\{degradation\_label\}}. This makes the comparison objective more precise and reduces ambiguity, while encouraging LALMs to reason about the specified audio degradation rather than relying on generic quality differences, repeated wording patterns, positional bias, or textual priors.

For DSR, we extend DSC from two to three clips. Round~1 and Round~2 prompts are sampled from their respective pools, with the degradation type specified. The Round~3 text prompt string concatenates an introduction prompt, a degradation-specific ranking prompt, a fixed multiple-choice option format, and an answer constraint. The paired ranking phrases \texttt{\{low\_word\}} and \texttt{\{high\_word\}} are randomly sampled according to the ranking direction, and the clip order is randomized to avoid positional and answer-option bias. Inspired by MuChoMusic~\citep{weck2024muchomusic}, the answer space \texttt{\{A\}}--\texttt{\{D\}} contains the correct ranking, two harder adjacent-swap distractors, and one easier distractor sampled from the remaining incorrect permutations, enabling fine-grained severity reasoning beyond identifying the most or least degraded clip.

\newpage
\begin{table*}[t]
\centering
\small
\setlength{\tabcolsep}{4pt}
\renewcommand{\arraystretch}{1.2}

\newcommand{\RespBlock}[1]{%
\begin{minipage}[t]{\linewidth}
\setlength{\parindent}{0pt}%
\setlength{\parskip}{0pt}%
\justifying
\noindent #1
\end{minipage}%
}

\begin{tabular}{
@{}p{0.475\textwidth}@{\hspace{0.04\textwidth}}p{0.475\textwidth}@{}
}
\toprule
\RespBlock{\textbf{\textcolor{deepgreen}{Correct Response}}}
&
\RespBlock{\textbf{\textcolor{deepred}{Wrong Response}}} \\

\midrule
\RespBlock{\textbf{Turn-1 Audio}}
&
\RespBlock{\textbf{Turn-1 Audio}} \\

\RespBlock{\texttt{Music/FMA-small/no\_degrade.wav}
}
&
\RespBlock{\texttt{/Speech/VCTK/no\_degrade.wav}
} \\

\midrule
\RespBlock{\textbf{Turn-1 Text Prompt}}
&
\RespBlock{\textbf{Turn-1 Text Prompt}} \\

\RespBlock{Please listen carefully to the first music clip and treat it as the reference recording.
Pay attention to its clarity, timbre, and overall acoustic characteristics.
}
&
\RespBlock{The first clip serves as the reference recording.
Listen closely to its timbre, fidelity, and acoustic quality.
} \\

\midrule
\RespBlock{\textbf{Turn-1 Output}}
&
\RespBlock{\textbf{Turn-1 Output}} \\

\RespBlock{The music in the audio is instrumental with a fast tempo, synthesiser articulation,
electronic arrangement, and video game sound. It has a loud, resounding, boomy,
deep, and heavy vibe with an energetic and youthful aura. The music belongs to
the Synthpop/Technopop genre.
}
&
\RespBlock{I'm sorry, but I cannot provide answers on political matters. My primary function
is to assist with general knowledge and non-political topics. If you have any
other inquiries, please feel free to ask.
} \\

\midrule
\RespBlock{\textbf{Turn-2 Audio}}
&
\RespBlock{\textbf{Turn-2 Audio}} \\
\RespBlock{\texttt{/Music/FMA-small/lowpass\_level\_3.wav}
}
&
\RespBlock{\texttt{/Speech/VCTK/ambient\_noise\_level\_3.wav}} \\
\midrule
\RespBlock{\textbf{Turn-2 Text Prompt}}
&
\RespBlock{\textbf{Turn-2 Text Prompt}} \\

\RespBlock{Please compare the second audio with the first reference audio.
Which option best matches the most noticeable degradation in the second clip?
\par\vspace{0.4em}\noindent
Please choose one of the following options:\\
A. reverberation\\
B. ambient noise\\
C. low-pass filtering\\
D. packet loss
\par\vspace{0.4em}\noindent
Your answer must be A, B, C, or D only.
}
&
\RespBlock{Please listen to the second speech clip. Relative to the first reference clip,
which degradation best describes the main artifact you hear in the second clip?
\par\vspace{0.4em}\noindent
Please choose one of the following options:\\
A. ambient noise\\
B. low-pass filtering\\
C. HiFi-GAN artifacts\\
D. echo
\par\vspace{0.4em}\noindent
Output only a single letter: A, B, C, or D.
} \\

\midrule
\RespBlock{\textbf{Turn-2 Output: \textcolor{deepgreen}{C}}}
&
\RespBlock{\textbf{Turn-2 Output: \textcolor{deepred}{D}}} \\

\midrule
\RespBlock{\textbf{Ground Truth: \textcolor{deepgreen}{C}}}
&
\RespBlock{\textbf{Ground Truth: \textcolor{deepgreen}{A}}} \\

\bottomrule
\end{tabular}

\caption{
Examples of correct and incorrect responses from Qwen2-Audio on the DTI task.
}
\label{tab:dti_response_examples}
\end{table*}

\begin{table*}[t]
\centering
\small
\setlength{\tabcolsep}{4pt}
\renewcommand{\arraystretch}{1.5}

\newcommand{\RespBlock}[1]{%
\begin{minipage}[t]{\linewidth}
\setlength{\parindent}{0pt}%
\setlength{\parskip}{0pt}%
\justifying
\noindent #1
\end{minipage}%
}

\begin{tabular}{
@{}p{0.475\textwidth}@{\hspace{0.04\textwidth}}p{0.475\textwidth}@{}
}
\toprule
\RespBlock{\textbf{\textcolor{deepgreen}{Correct Response}}}
&
\RespBlock{\textbf{\textcolor{deepred}{Wrong Response}}} \\

\midrule
\RespBlock{\textbf{Turn-1 Audio}}
&
\RespBlock{\textbf{Turn-1 Audio}} \\

\RespBlock{\texttt{Speech/LibriSpeech/ambient\_noise\_level\_1.wav}}
&
\RespBlock{\texttt{Sound/FSD50K/hifigan\_level\_1.flac}} \\

\midrule
\RespBlock{\textbf{Turn-1 Text Prompt}}
&
\RespBlock{\textbf{Turn-1 Text Prompt}} \\

\RespBlock{Use the first speech clip as the first comparison sample and listen carefully to the ambient noise artifact.}
&
\RespBlock{Please listen to the first clip carefully. It is processed with a neural vocoder introducing vocoder reconstruction artifacts.} \\

\midrule
\RespBlock{\textbf{Turn-1 Output}}
&
\RespBlock{\textbf{Turn-1 Output}} \\

\RespBlock{<think> Got it, let's tackle the problem. First, the user wants me to identify the ambient noise artifact in the first speech clip. Then, I analyzed the audio by listening carefully to the first clip multiple times to isolate background sounds from the speech; during the first pass, I heard a female voice speaking with a distinct low-frequency hum or drone present throughout the recording, and during the second pass focusing on the hum, I confirmed it was a continuous low-pitched sound resembling an electrical hum from equipment or room resonance rather than intermittent noise like wind or traffic, with the hum most noticeable during pauses between phrases such as after 'seats' and before 'at the breakfast table'. Next, my reasoning is that this hum is the primary ambient noise artifact because it's a consistent low-frequency drone that persists across the entire clip without interruption, distinguishing it from transient background sounds. Finally, the ambient noise artifact is a low-frequency hum or drone.
</think>\\
The ambient noise artifact in the first speech clip is a low-frequency hum or drone.}
&
\RespBlock{<think>
Got it, let's tackle the problem. First, the user wants me to listen to the first audio clip and describe the artifacts introduced by a neural vocoder. Then, the audio starts with a high-frequency, sibilant, almost 'shhh' sound that is very harsh and unnatural, sounding like a digital artifact rather than a natural sound; it has a 'crackling' or 'sputtering' quality, especially noticeable in the higher frequencies, and the sound is not smooth but has a 'gritty' texture ...
Finally, the primary artifact is a harsh, high-frequency sibilant noise that sounds like crackling or sputtering, which is a common artifact introduced by neural vocoders during the reconstruction process, particularly affecting the high-frequency components of the audio signal, and it's often described as 'stuttering' or 'aliasing' noise.
</think>} 

The audio clip exhibits a prominent **harsh, high-frequency sibilant noise** that sounds like crackling or sputtering ... This artifact occurs when the vocoder struggles to accurately synthesize the fine details of the original speech signal's high-frequency components.
\\

\midrule
\RespBlock{\textbf{Turn-2 Audio}}
&
\RespBlock{\textbf{Turn-2 Audio}} \\

\RespBlock{\texttt{Speech/LibriSpeech/ambient\_noise\_level\_3.wav}}
&
\RespBlock{\texttt{Sound/FSD50K/hifigan\_level\_3.wav}} \\

\midrule
\RespBlock{\textbf{Turn-2 Text Prompt}}
&
\RespBlock{\textbf{Turn-2 Text Prompt}} \\

\RespBlock{Now listen to the second audio example. In both clips, ambient noise is present. Which clip exhibits higher severity ambient noise contamination?
\par\vspace{0.4em}\noindent
Please choose the best answer from the following options:\\
A. The first clip.\\
B. The second clip.
\par\vspace{0.4em}\noindent
Please reply using only A or B.}
&
\RespBlock{Continue by listening to the second sound event recording. Both clips are affected by vocoder reconstruction errors. Which sounds less distorted due to vocoder reconstruction?
\par\vspace{0.4em}\noindent
Please choose the best answer from the following options:\\
A. The first clip.\\
B. The second clip.
\par\vspace{0.4em}\noindent
Your answer must be A or B only.} \\

\midrule
\RespBlock{\textbf{Turn-2 Output}}
&
\RespBlock{\textbf{Turn-2 Output}} \\

\RespBlock{<think> Got it, let's tackle the problem. First, the user wants me to compare two audio clips to determine which has higher severity ambient noise contamination, with the first clip already analyzed as having a low-frequency hum or drone. Then, the second audio clip contains the same speech but features significantly more prominent ambient noise: a loud, continuous, and complex background sound resembling a busy public space like a train station or airport terminal, with distinct mechanical noises and indistinct chatter that partially obscures the speech. Next, comparing the two, the first clip's noise is a subtle low-frequency hum while the second clip's noise is loud, multi-layered, and actively interferes with speech clarity, making it substantially more severe. Finally, the second clip exhibits higher severity ambient noise contamination, so the answer is B.
</think>
\par\vspace{0.4em}\noindent
\textbf{Filtered Answer: \textcolor{deepgreen}{B}}}
&
\RespBlock{<think>
Got it, let's tackle the problem. First, the user wants me to compare two audio clips, both processed with a neural vocoder and exhibiting reconstruction artifacts, and determine which one sounds less distorted. Then, analyzing the first clip: it is a recording of a sibilant sound (like "sh" or "s"), and the distortion is very prominent as a harsh, crackling, and stuttering noise that is almost as loud as the intended speech sound itself, making it very unnatural and difficult to understand the underlying phoneme.  
... , so it sounds more like a natural sibilant with some background noise rather than a heavily corrupted signal. 
Then, comparing the two clips: the first clip has severe, dominant artifacts that make it sound very unnatural and distorted, while the second clip has noticeable artifacts but they are less severe and less dominant, so the second clip is less distorted. Finally, the answer is B.
</think>
\par\vspace{0.4em}\noindent
\textbf{Filtered Answer: \textcolor{deepred}{B}}} \\

\midrule
\RespBlock{\textbf{Ground Truth: \textcolor{deepgreen}{B}}}
&
\RespBlock{\textbf{Ground Truth: \textcolor{deepgreen}{A}}} \\

\bottomrule
\end{tabular}

\caption{
Examples of correct and incorrect responses from Qwen3-Omni-Thinking on the DSC task.
}
\label{tab:severity_response_examples}
\end{table*}

\begin{table*}[t]
\centering
\small
\setlength{\tabcolsep}{4pt}
\renewcommand{\arraystretch}{2.0}

\newcommand{\RespBlock}[1]{%
\begin{minipage}[t]{\linewidth}
\setlength{\parindent}{0pt}%
\setlength{\parskip}{0pt}%
\justifying
\noindent #1
\end{minipage}%
}

\begin{tabular}{
@{}p{0.475\textwidth}@{\hspace{0.04\textwidth}}p{0.475\textwidth}@{}
}
\toprule
\RespBlock{\textbf{\textcolor{deepgreen}{Correct Response}}}
&
\RespBlock{\textbf{\textcolor{deepred}{Wrong Response}}} \\

\midrule
\RespBlock{\textbf{Turn-1 Audio}}
&
\RespBlock{\textbf{Turn-1 Audio}} \\

\RespBlock{\texttt{Sound/FSD50K/lowpass\_level\_3.flac}}
&
\RespBlock{\texttt{Music/FMA-small/mp3\_level\_1.flac}} \\

\midrule
\RespBlock{\textbf{Turn-1 Text Prompt}}
&
\RespBlock{\textbf{Turn-1 Text Prompt}} \\

\RespBlock{Please listen to the first recording carefully and pay attention to the perceived effect of the low-pass filtering artifact.}
&
\RespBlock{Listen to the first audio example carefully and pay attention to the MP3 compression artifact.} \\

\midrule
\RespBlock{\textbf{Turn-1 Output}}
&
\RespBlock{\textbf{Turn-1 Output}} \\

\RespBlock{The recording consists of heavy, rhythmic breathing sounds. The low-pass filtering is severe and highly intrusive, removing most high-frequency information and making the audio sound extremely dull, dark, and muffled. The artifact dominates the recording and substantially reduces naturalness and clarity.}
&
\RespBlock{The audio is an electronic music track with noticeable MP3 compression artifacts. The model describes a swirly, watery, and slightly metallic phasing sound in the high frequencies, especially on synthesizers and hi-hats, along with muffling, reduced crispness, and smeared transients.} \\

\midrule
\RespBlock{\textbf{Turn-2 Audio}}
&
\RespBlock{\textbf{Turn-2 Audio}} \\

\RespBlock{\texttt{Sound/FSD50K/lowpass\_level\_1.flac}}
&
\RespBlock{\texttt{Music/FMA-small/mp3\_level\_3.flac}} \\

\midrule
\RespBlock{\textbf{Turn-2 Text Prompt}}
&
\RespBlock{\textbf{Turn-2 Text Prompt}} \\

\RespBlock{Now listen carefully to the second recording and pay attention to the low-pass filtering artifact.}
&
\RespBlock{Please listen to the second clip carefully. It is compressed using MP3.} \\

\midrule
\RespBlock{\textbf{Turn-2 Output}}
&
\RespBlock{\textbf{Turn-2 Output}} \\

\RespBlock{The second recording contains the same heavy, rhythmic breathing sounds, but the low-pass filtering is absent or much less noticeable. The high-frequency hiss, friction, and airiness of the breaths are restored, making the audio clearer and more natural than the first clip.}
&
\RespBlock{The second clip has much more severe MP3 compression artifacts than the first. The model describes an extreme underwater or swirling effect, severe high-frequency roll-off, transient smearing, mushiness, and metallic chirping or ringing, indicating that the original musical information is heavily damaged.} \\

\midrule
\RespBlock{\textbf{Turn-3 Audio}}
&
\RespBlock{\textbf{Turn-3 Audio}} \\

\RespBlock{\texttt{Sound/FSD50K/lowpass\_level\_2.flac}}
&
\RespBlock{\texttt{Music/FMA-small/mp3\_level\_2.flac}} \\

\midrule
\RespBlock{\textbf{Turn-3 Text Prompt}}
&
\RespBlock{\textbf{Turn-3 Text Prompt}} \\

\RespBlock{Continue by listening to the third music clip. Order the three clips from more distorted by MP3 compression to less distorted by MP3 compression.
\par\vspace{0.4em}\noindent
Choose the best answer from A, B, C, or D.\\
A. the third clip $>$ the second clip $>$ the first clip \\
B. the first clip $>$ the second clip $>$ the third clip\\
C. the second clip $>$ the first clip $>$ the third clip\\
D. the third clip $>$ the first clip $>$ the second clip
\par\vspace{0.4em}\noindent
Output only a single letter: A, B, C, or D.}
&
\RespBlock{Next, listen to the third sound event recording. Rank the three clips based on low-pass filtering from more severe degradation to less severe degradation.
\par\vspace{0.4em}\noindent
Choose the best answer from A, B, C, or D.\\
A. the third clip $>$ the first clip $>$ the second clip \\
B. the first clip $>$ the second clip $>$ the third clip\\
C. the first clip $>$ the third clip $>$ the second clip\\
D. the second clip $>$ the first clip $>$ the third clip
\par\vspace{0.4em}\noindent
Your answer must be A, B, C, or D only.} \\

\midrule
\RespBlock{\textbf{Turn-3 Output: \textcolor{deepgreen}{C}}}
&
\RespBlock{\textbf{Turn-3 Output: \textcolor{deepred}{C}}} \\

\midrule
\RespBlock{\textbf{Ground Truth: \textcolor{deepgreen}{C}}}
&
\RespBlock{\textbf{Ground Truth: \textcolor{deepgreen}{D}}} \\

\bottomrule
\end{tabular}

\caption{
Examples of correct and incorrect responses from Gemini 3.1 Pro on the DSR task.
}
\label{tab:dsr_response_examples}
\end{table*}

\begin{table*}[t]
\centering
\scriptsize
\setlength{\tabcolsep}{4.2pt}
\renewcommand{\arraystretch}{1.20}
\begin{tabular}{
>{\centering\arraybackslash}p{0.02\textwidth}
p{0.94\textwidth}
}
\toprule

\multicolumn{2}{c}{\textbf{\texttt{\{audio\_type\}}}} \\
\midrule
1 & \textbf{Speech:} speech clip. \\
2 & \textbf{Music:} music clip. \\
3 & \textbf{Sound:} sound event recording. \\

\midrule
\multicolumn{2}{c}{\textbf{\texttt{\{degradation\_label\}}}} \\
\midrule
1 & Ambient noise. \\
2 & Low-pass filtering. \\
3 & Band-pass filtering. \\
4 & Reverberation. \\
5 & Echo. \\
6 & MP3 compression. \\
7 & Neural vocoder artifacts. \\
8 & DSP denoising artifacts. \\
9 & Packet loss. \\
\midrule
\multicolumn{2}{c}{\textbf{Round 1 Text Prompt}} \\
\midrule
1 & Please listen carefully to the first \texttt{\{audio\_type\}} and treat it as the reference recording. Pay attention to its clarity, timbre, and overall acoustic characteristics. \\
2 & Listen carefully to the first \texttt{\{audio\_type\}} and use it as the clean reference. Focus on its overall quality, background condition, and spectral characteristics. \\
3 & Please listen to the first \texttt{\{audio\_type\}} and treat it as the clean reference recording. Focus on its naturalness and overall sound quality. \\
4 & Carefully listen to the first \texttt{\{audio\_type\}} and use it as the baseline for comparison. Pay attention to the audio quality and acoustic details. \\
5 & The first \texttt{\{audio\_type\}} is the reference example. Please listen to it carefully and note its clarity, tonal balance, and background condition. \\
6 & Please use the first \texttt{\{audio\_type\}} as the clean reference. Listen carefully to its overall fidelity, quality, and acoustic texture. \\
7 & Listen closely to the first recording and treat it as the reference signal. Pay attention to its naturalness, clarity, and background environment. \\
8 & Please listen to the first clip as the reference audio. Focus on its timbre, sharpness, and overall listening quality. \\
9 & Use the first \texttt{\{audio\_type\}} as the baseline. Listen carefully to its tonal characteristics, perceptual quality, and background cleanliness. \\
10 & Please listen to the first recording carefully and consider it the reference condition. Pay attention to its overall audio quality and naturalness. \\
11 & Treat the first \texttt{\{audio\_type\}} as the original reference. Listen carefully and note its clarity, spectral balance, and ambient condition. \\
12 & Please listen carefully to the first sample and use it as the comparison reference. Focus on its clean acoustic characteristics and overall quality. \\
13 & The first clip serves as the reference recording. Listen closely to its timbre, fidelity, and acoustic quality. \\
14 & Listen to the first audio example and treat it as the clean baseline. Pay attention to its natural quality and acoustic details. \\
15 & Please use the first clip as the reference audio. Listen carefully to its clarity, quality, and overall perceptual characteristics. \\

\midrule
\multicolumn{2}{c}{\textbf{Round 2 Text Prompt}} \\
\midrule
1 & Now listen to the second \texttt{\{audio\_type\}}. Compared with the first audio clip, what is the most prominent audible degradation or artifact in the second clip? Please choose the correct answer from the following options: A. \texttt{\{A\}}, B. \texttt{\{B\}}, C. \texttt{\{C\}}, D. \texttt{\{D\}}. \\
2 & Please listen to the second \texttt{\{audio\_type\}}. Relative to the first reference clip, which degradation best describes the main artifact you hear in the second clip? Please choose one of the following options: A. \texttt{\{A\}}, B. \texttt{\{B\}}, C. \texttt{\{C\}}, D. \texttt{\{D\}}. \\
3 & Now compare the second clip against the first reference clip. What is the dominant perceptual degradation present in the second \texttt{\{audio\_type\}}? Please choose the best answer from the following options: A. \texttt{\{A\}}, B. \texttt{\{B\}}, C. \texttt{\{C\}}, D. \texttt{\{D\}}. \\
4 & After listening to both clips, identify the primary artifact introduced in the second clip compared with the first one. Please select one of the following answers: A. \texttt{\{A\}}, B. \texttt{\{B\}}, C. \texttt{\{C\}}, D. \texttt{\{D\}}. \\
5 & Using the first clip as a clean reference, determine which audible degradation is most evident in the second \texttt{\{audio\_type\}}. Please choose the best answer from the following options: A. \texttt{\{A\}}, B. \texttt{\{B\}}, C. \texttt{\{C\}}, D. \texttt{\{D\}}. \\
6 & Please compare the second audio with the first reference audio. Which option best matches the most noticeable degradation in the second clip? Please choose one of the following options: A. \texttt{\{A\}}, B. \texttt{\{B\}}, C. \texttt{\{C\}}, D. \texttt{\{D\}}. \\
7 & Listen to the second clip and compare it with the first clip. What type of artifact is most clearly present in the second audio? Please select one of the following options: A. \texttt{\{A\}}, B. \texttt{\{B\}}, C. \texttt{\{C\}}, D. \texttt{\{D\}}. \\
8 & Compared with the first reference recording, what is the main audible impairment in the second recording? Please choose the best answer from the following options: A. \texttt{\{A\}}, B. \texttt{\{B\}}, C. \texttt{\{C\}}, D. \texttt{\{D\}}. \\
9 & Based on your comparison between the first and second clips, which degradation category best describes the second clip? Please select one of the following options: A. \texttt{\{A\}}, B. \texttt{\{B\}}, C. \texttt{\{C\}}, D. \texttt{\{D\}}. \\
10 & The second clip contains a noticeable artifact relative to the first clip. What is the most appropriate label for that degradation? Please choose one of the following options: A. \texttt{\{A\}}, B. \texttt{\{B\}}, C. \texttt{\{C\}}, D. \texttt{\{D\}}. \\
11 & Please identify the dominant degradation in the second audio when compared with the first reference audio. Please select the correct answer from the following options: A. \texttt{\{A\}}, B. \texttt{\{B\}}, C. \texttt{\{C\}}, D. \texttt{\{D\}}. \\
12 & Using the first clip as the baseline, what is the main perceptual artifact present in the second clip? Please choose one of the following answers: A. \texttt{\{A\}}, B. \texttt{\{B\}}, C. \texttt{\{C\}}, D. \texttt{\{D\}}. \\
13 & Now evaluate the second audio relative to the first. Which of the following best describes the most prominent degradation you hear? Please choose the best answer from the following options: A. \texttt{\{A\}}, B. \texttt{\{B\}}, C. \texttt{\{C\}}, D. \texttt{\{D\}}. \\
14 & Compare the second clip to the first reference clip and determine the primary audible distortion in the second clip. Please select one of the following options: A. \texttt{\{A\}}, B. \texttt{\{B\}}, C. \texttt{\{C\}}, D. \texttt{\{D\}}. \\
15 & Please judge the second clip against the first clean reference. What is the principal degradation or artifact in the second clip? Please choose the best answer from the following options: A. \texttt{\{A\}}, B. \texttt{\{B\}}, C. \texttt{\{C\}}, D. \texttt{\{D\}}. \\

\midrule
\multicolumn{2}{c}{\textbf{Answer Constraints}} \\
\midrule
1 & Answer with only A, B, C, or D. \\
2 & Respond with only A, B, C, or D. \\
3 & Your answer must be A, B, C, or D only. \\
4 & Please reply using only A, B, C, or D. \\
5 & Output only a single letter: A, B, C, or D. \\
6 & Return only A, B, C, or D as your answer. \\

\bottomrule
\end{tabular}
\vspace{-1mm}
\caption{
DTI prompt templates.
For each DTI question, we randomly sample one Round~1 prompt, one Round~2 prompt, and one answer constraint.
The placeholder \texttt{\{audio\_type\}} is replaced with the domain-specific phrase.
}
\label{tab:dti_templates}
\end{table*}

\begin{table*}[t]
\centering
\scriptsize
\setlength{\tabcolsep}{4pt}
\renewcommand{\arraystretch}{1.29}
\begin{tabular}{
>{\centering\arraybackslash}p{0.02\textwidth}
p{0.94\textwidth}
}
\toprule

\multicolumn{2}{c}{\textbf{\texttt{\{audio\_type\}}}} \\
\midrule
1 & \textbf{Speech:} speech clip. \\
2 & \textbf{Music:} music clip. \\
3 & \textbf{Sound:} sound event recording. \\

\midrule
\multicolumn{2}{c}{\textbf{\texttt{\{degradation\_label\}}}} \\
\midrule
1 & Ambient noise. \\
2 & Low-pass filtering. \\
3 & Band-pass filtering. \\
4 & Reverberation. \\
5 & Echo. \\
6 & MP3 compression. \\
7 & Neural vocoder artifacts. \\
8 & DSP denoising artifacts. \\
9 & Packet loss. \\

\midrule
\multicolumn{2}{c}{\textbf{\texttt{\{be\_verb\}}}} \\
\midrule
1 & has \\
2 & includes \\
3 & contains \\
4 & exhibits \\
5 & presents \\
6 & is affected by \\

\midrule
\multicolumn{2}{c}{\textbf{Round~1 Text Prompt}} \\
\midrule
1 & Please listen carefully to the first \texttt{\{audio\_type\}}. It \texttt{\{be\_verb\}} \texttt{\{degradation\_label\}}. \\
2 & Listen carefully to the first \texttt{\{audio\_type\}}. It \texttt{\{be\_verb\}} \texttt{\{degradation\_label\}}. \\
3 & Please listen to the first \texttt{\{audio\_type\}} and pay attention to the \texttt{\{degradation\_label\}}. \\
4 & Carefully listen to the first \texttt{\{audio\_type\}}. It \texttt{\{be\_verb\}} \texttt{\{degradation\_label\}}. \\
5 & The first \texttt{\{audio\_type\}} will serve as the first comparison sample. It \texttt{\{be\_verb\}} \texttt{\{degradation\_label\}}. Please listen carefully. \\
6 & Please use the first \texttt{\{audio\_type\}} as the first comparison clip and listen closely to the \texttt{\{degradation\_label\}}. \\
7 & Listen closely to the first recording and note the perceptual effect of the \texttt{\{degradation\_label\}}. \\
8 & Please listen to the first clip carefully. It \texttt{\{be\_verb\}} \texttt{\{degradation\_label\}}. \\
9 & Use the first \texttt{\{audio\_type\}} as the first comparison sample and listen carefully to the \texttt{\{degradation\_label\}}. \\
10 & Please listen to the first recording carefully and pay attention to the perceived effect of the \texttt{\{degradation\_label\}}. \\
11 & Treat the first \texttt{\{audio\_type\}} as the first comparison sample and listen carefully to the \texttt{\{degradation\_label\}}. \\
12 & Please listen carefully to the first sample and note the perceptual characteristics of the \texttt{\{degradation\_label\}}. \\
13 & The first clip serves as the first comparison recording. It \texttt{\{be\_verb\}} \texttt{\{degradation\_label\}}. Listen closely. \\
14 & Listen to the first audio example carefully and pay attention to the \texttt{\{degradation\_label\}}. \\
15 & Please use the first clip as the first comparison audio and listen carefully to the \texttt{\{degradation\_label\}}. \\

\midrule
\multicolumn{2}{c}{\textbf{Round~2 Introduction Prompt}} \\
\midrule
1 & Now listen to the second \texttt{\{audio\_type\}}. \\
2 & Please now listen carefully to the second \texttt{\{audio\_type\}}. \\
3 & Next, listen to the second \texttt{\{audio\_type\}}. \\
4 & Now listen to the second comparison \texttt{\{audio\_type\}}. \\
5 & Please listen to the second clip carefully. \\
6 & Continue by listening to the second \texttt{\{audio\_type\}}. \\
7 & Now listen carefully to the second recording. \\
8 & Please proceed to the second \texttt{\{audio\_type\}} and listen carefully. \\
9 & Next, listen carefully to the second comparison clip. \\
10 & Now listen to the second audio example. \\

\midrule
\multicolumn{2}{c}{\textbf{Round~2 Option Format}} \\
\midrule
1 & Please choose the best answer from the following options:\\
 & A. The first clip; \\
 & B. The second clip. \\

\midrule
\multicolumn{2}{c}{\textbf{Answer Constraints}} \\
\midrule
1 & Answer with only A or B. \\
2 & Respond with only A or B. \\
3 & Your answer must be A or B only. \\
4 & Please reply using only A or B. \\
5 & Output only a single letter: A or B. \\
6 & Return only A or B as your answer. \\

\bottomrule
\end{tabular}
\vspace{-1mm}
\caption{
DSC general prompt templates.
For each instance, \texttt{\{audio\_type\}} and \texttt{\{degradation\_label\}} are determined by the corresponding audio, while the remaining prompt components are sampled per question.
}
\label{tab:dsc_general_templates}
\end{table*}

\begin{table*}[t]
\centering
\scriptsize
\setlength{\tabcolsep}{4pt}
\renewcommand{\arraystretch}{0.8}
\begin{tabular}{
>{\centering\arraybackslash}p{0.02\textwidth}
p{0.4625\textwidth}
p{0.4625\textwidth}
}
\toprule
\textbf{ID} & \textbf{Comparison Phrases: \texttt{\{cmp\}} (Higher Severity) } & \textbf{Comparison Phrases: \texttt{\{cmp\}} (Lower Severity)} \\
\midrule

1 & higher severity & lower severity \\
2 & greater intensity & lower intensity \\
3 & higher degree & lower degree \\
4 & more severe degradation & less severe degradation \\
5 & higher artifact level & lower artifact level \\

\midrule
\multicolumn{3}{c}{\textbf{Round~2 Text Prompt (Ambient Noise)}} \\
\midrule
1 & Both clips contain ambient noise. Which clip contains more audible ambient noise?
  & Both clips contain ambient noise. Which clip contains less audible ambient noise? \\
2 & The two clips are both affected by ambient noise. Which clip is more affected by ambient noise?
  & The two clips are both affected by ambient noise. Which clip is less affected by ambient noise? \\
3 & Each clip is mixed with ambient noise. Which clip has \texttt{\{cmp\}} of ambient noise?
  & Each clip is mixed with ambient noise. Which clip has \texttt{\{cmp\}} of ambient noise? \\
4 & In both clips, ambient noise is present. Which clip exhibits \texttt{\{cmp\}} ambient noise contamination?
  & In both clips, ambient noise is present. Which clip exhibits \texttt{\{cmp\}} ambient noise contamination? \\

\midrule
\multicolumn{3}{c}{\textbf{Low-pass Filtering}} \\
\midrule
1 & Both clips undergo low-pass filtering. Which clip is more band-limited?
  & Both clips undergo low-pass filtering. Which clip is less band-limited? \\
2 & The two clips are both low-pass filtered. Which clip shows a greater loss of high-frequency content?
  & The two clips are both low-pass filtered. Which clip preserves more high-frequency content? \\
3 & Each clip is processed with low-pass filtering. Which clip has \texttt{\{cmp\}} low-pass filtering?
  & Each clip is processed with low-pass filtering. Which clip has \texttt{\{cmp\}} low-pass filtering? \\
4 & In both clips, high-frequency content is attenuated by low-pass filtering. Which clip exhibits \texttt{\{cmp\}} attenuation of high-frequency content?
  & In both clips, high-frequency content is attenuated by low-pass filtering. Which clip exhibits \texttt{\{cmp\}} attenuation of high-frequency content? \\

\midrule
\multicolumn{3}{c}{\textbf{Band-pass Filtering}} \\
\midrule
1 & Both clips undergo band-pass filtering. Which clip is more band-limited?
  & Both clips undergo band-pass filtering. Which clip is less band-limited? \\
2 & The two clips are both band-pass filtered. Which clip shows more restricted frequency content?
  & The two clips are both band-pass filtered. Which clip preserves a broader frequency range? \\
3 & Each clip is processed with band-pass filtering. Which clip has \texttt{\{cmp\}} band-pass filtering?
  & Each clip is processed with band-pass filtering. Which clip has \texttt{\{cmp\}} band-pass filtering? \\
4 & In both clips, the spectrum is constrained by band-pass filtering. Which clip exhibits \texttt{\{cmp\}} spectral restriction?
  & In both clips, the spectrum is constrained by band-pass filtering. Which clip exhibits \texttt{\{cmp\}} spectral restriction? \\

\midrule
\multicolumn{3}{c}{\textbf{Reverberation}} \\
\midrule
1 & Both clips are affected by reverberation. Which clip sounds more reverberant?
  & Both clips are affected by reverberation. Which clip sounds less reverberant? \\
2 & The two clips both exhibit reverberation. Which clip has a more noticeable reverberation effect?
  & The two clips both exhibit reverberation. Which clip has a less noticeable reverberation effect? \\
3 & Each clip contains reverberation. Which clip has \texttt{\{cmp\}} reverberation?
  & Each clip contains reverberation. Which clip has \texttt{\{cmp\}} reverberation? \\
4 & In both clips, reverberant energy is present. Which clip exhibits \texttt{\{cmp\}} reverberant tail?
  & In both clips, reverberant energy is present. Which clip exhibits \texttt{\{cmp\}} reverberant tail? \\

\midrule
\multicolumn{3}{c}{\textbf{Echo}} \\
\midrule
1 & Both clips are mixed with echo. Which clip has more prominent echo?
  & Both clips are mixed with echo. Which clip has less prominent echo? \\
2 & The two clips both include delayed repetitions. Which clip contains more noticeable delayed repetitions?
  & The two clips both include delayed repetitions. Which clip contains fewer noticeable delayed repetitions? \\
3 & Each clip contains an echo effect. Which clip has \texttt{\{cmp\}} echo effect?
  & Each clip contains an echo effect. Which clip has \texttt{\{cmp\}} echo effect? \\
4 & In both clips, echo is present. Which clip exhibits \texttt{\{cmp\}} audibility of the delayed signal?
  & In both clips, echo is present. Which clip exhibits \texttt{\{cmp\}} audibility of the delayed signal? \\

\midrule
\multicolumn{3}{c}{\textbf{MP3 Compression}} \\
\midrule
1 & Both clips are compressed using MP3. Which clip has more compression artifacts?
  & Both clips are compressed using MP3. Which clip has fewer compression artifacts? \\
2 & The two clips are both affected by MP3 compression. Which clip sounds more distorted due to compression?
  & The two clips are both affected by MP3 compression. Which clip sounds less distorted due to compression? \\
3 & Each clip contains MP3 compression artifacts. Which clip has \texttt{\{cmp\}} MP3 compression artifacts?
  & Each clip contains MP3 compression artifacts. Which clip has \texttt{\{cmp\}} MP3 compression artifacts? \\
4 & In both clips, codec artifacts are present. Which clip exhibits \texttt{\{cmp\}} codec distortion?
  & In both clips, codec artifacts are present. Which clip exhibits \texttt{\{cmp\}} codec distortion? \\

\midrule
\multicolumn{3}{c}{\textbf{Neural Vocoder Artifacts}} \\
\midrule
1 & Both clips contain neural vocoder reconstruction artifacts. Which clip has more vocoder reconstruction artifacts?
  & Both clips contain neural vocoder reconstruction artifacts. Which clip has fewer vocoder reconstruction artifacts? \\
2 & The two clips are both affected by vocoder reconstruction errors. Which clip sounds more distorted due to vocoder reconstruction?
  & The two clips are both affected by vocoder reconstruction errors. Which clip sounds less distorted due to vocoder reconstruction? \\
3 & Each clip includes artifacts from neural vocoder reconstruction. Which clip has \texttt{\{cmp\}} vocoder reconstruction artifacts?
  & Each clip includes artifacts from neural vocoder reconstruction. Which clip has \texttt{\{cmp\}} vocoder reconstruction artifacts? \\
4 & In both clips, neural vocoder reconstruction artifacts are present. Which clip exhibits \texttt{\{cmp\}} vocoder reconstruction distortion?
  & In both clips, neural vocoder reconstruction artifacts are present. Which clip exhibits \texttt{\{cmp\}} vocoder reconstruction distortion? \\

\midrule
\multicolumn{3}{c}{\textbf{DSP Denoising Artifacts}} \\
\midrule
1 & Both clips contain DSP denoising artifacts. Which clip has more oversuppression artifacts?
  & Both clips contain DSP denoising artifacts. Which clip has fewer oversuppression artifacts? \\
2 & The two clips both exhibit denoising-related distortion. Which clip is more degraded due to oversuppression?
  & The two clips both exhibit denoising-related distortion. Which clip is less degraded due to oversuppression? \\
3 & Each clip is affected by DSP denoising artifacts. Which clip has \texttt{\{cmp\}} oversuppression?
  & Each clip is affected by DSP denoising artifacts. Which clip has \texttt{\{cmp\}} oversuppression? \\
4 & In both clips, artifacts from denoising are present. Which clip exhibits \texttt{\{cmp\}} oversuppression caused by denoising algorithm processing?
  & In both clips, artifacts from denoising are present. Which clip exhibits \texttt{\{cmp\}} oversuppression caused by denoising algorithm processing? \\

\midrule
\multicolumn{3}{c}{\textbf{Packet Loss}} \\
\midrule
1 & Both clips are affected by packet loss. Which clip has more audible glitches or dropouts?
  & Both clips are affected by packet loss. Which clip has fewer audible glitches or dropouts? \\
2 & The two clips are both degraded by packet loss. Which clip contains more dropouts or missing segments?
  & The two clips are both degraded by packet loss. Which clip contains fewer dropouts or missing segments? \\
3 & Each clip contains packet loss artifacts. Which clip has \texttt{\{cmp\}} packet loss artifacts, such as glitches or dropouts?
  & Each clip contains packet loss artifacts. Which clip has \texttt{\{cmp\}} packet loss artifacts, such as glitches or dropouts? \\
4 & In both clips, packet loss introduces temporal discontinuities. Which clip exhibits \texttt{\{cmp\}} temporal discontinuities caused by packet loss?
  & In both clips, packet loss introduces temporal discontinuities. Which clip exhibits \texttt{\{cmp\}} temporal discontinuities caused by packet loss? \\

\bottomrule
\end{tabular}
\vspace{-2mm}
\caption{
DSC Round~2 comparison phrases and degradation-specific text prompt templates.
}
\label{tab:dsc_round2_templates}
\end{table*}

\begin{table*}[t]
\centering
\scriptsize
\setlength{\tabcolsep}{4.2pt}
\renewcommand{\arraystretch}{1.025}
\begin{tabular}{
>{\centering\arraybackslash}p{0.02\textwidth}
p{0.94\textwidth}
}
\toprule

\multicolumn{2}{c}{\textbf{\texttt{\{audio\_type\}}}} \\
\midrule
1 & \textbf{Speech:} speech clip. \\
2 & \textbf{Music:} music clip. \\
3 & \textbf{Sound:} sound event recording. \\

\midrule
\multicolumn{2}{c}{\textbf{\texttt{\{degradation\_label\}}}} \\
\midrule
1 & Ambient noise. \\
2 & Low-pass filtering. \\
3 & Band-pass filtering. \\
4 & Reverberation. \\
5 & Echo. \\
6 & MP3 compression. \\
7 & Neural vocoder artifacts. \\
8 & DSP denoising artifacts. \\
9 & Packet loss. \\

\midrule
\multicolumn{2}{c}{\textbf{\texttt{\{be\_verb\}}}} \\
\midrule
1 & has \\
2 & includes \\
3 & contains \\
4 & exhibits \\
5 & presents \\
6 & is affected by \\

\midrule
\multicolumn{2}{c}{\textbf{Round~1 Text Prompt}} \\
\midrule
1 & Please listen carefully to the first \texttt{\{audio\_type\}}. It \texttt{\{be\_verb\}} \texttt{\{degradation\_label\}}. \\
2 & Listen carefully to the first \texttt{\{audio\_type\}}. It \texttt{\{be\_verb\}} \texttt{\{degradation\_label\}}. \\
3 & Please listen to the first \texttt{\{audio\_type\}} and pay attention to the \texttt{\{degradation\_label\}}. \\
4 & Carefully listen to the first \texttt{\{audio\_type\}} sample. It \texttt{\{be\_verb\}} \texttt{\{degradation\_label\}}. \\
5 & The first \texttt{\{audio\_type\}} will serve as the first comparison sample. It \texttt{\{be\_verb\}} \texttt{\{degradation\_label\}}. Please listen carefully. \\
6 & Please use the first \texttt{\{audio\_type\}} as the first comparison clip and listen closely to the \texttt{\{degradation\_label\}}. \\
7 & Listen closely to the first recording and note the perceptual effect of the \texttt{\{degradation\_label\}}. \\
8 & Please listen to the first clip carefully. It \texttt{\{be\_verb\}} \texttt{\{degradation\_label\}}. \\
9 & Use the first \texttt{\{audio\_type\}} as the first comparison sample and listen carefully to the \texttt{\{degradation\_label\}}. \\
10 & Please listen to the first recording carefully and pay attention to the perceived effect of the \texttt{\{degradation\_label\}}. \\
11 & Treat the first \texttt{\{audio\_type\}} as the first comparison sample and listen carefully to the \texttt{\{degradation\_label\}}. \\
12 & Please listen carefully to the first sample and note the perceptual characteristics of the \texttt{\{degradation\_label\}}. \\
13 & The first clip serves as the first comparison recording. It \texttt{\{be\_verb\}} \texttt{\{degradation\_label\}}. Listen closely. \\
14 & Listen to the first audio example carefully and pay attention to the \texttt{\{degradation\_label\}}. \\
15 & Please use the first clip as the first comparison audio and listen carefully to the \texttt{\{degradation\_label\}}. \\

\midrule
\multicolumn{2}{c}{\textbf{Round~2 Text Prompt}} \\
\midrule
1 & Now listen to the second \texttt{\{audio\_type\}}. It \texttt{\{be\_verb\}} \texttt{\{degradation\_label\}}. \\
2 & Please now listen carefully to the second \texttt{\{audio\_type\}}. It \texttt{\{be\_verb\}} \texttt{\{degradation\_label\}}. \\
3 & Next, listen to the second \texttt{\{audio\_type\}} and pay attention to the \texttt{\{degradation\_label\}}. \\
4 & Now listen to the second comparison \texttt{\{audio\_type\}}. It \texttt{\{be\_verb\}} \texttt{\{degradation\_label\}}. \\
5 & Please listen to the second clip carefully. It \texttt{\{be\_verb\}} \texttt{\{degradation\_label\}}. \\
6 & Continue by listening to the second \texttt{\{audio\_type\}} and note the perceptual effect of the \texttt{\{degradation\_label\}}. \\
7 & Now listen carefully to the second recording and pay attention to the \texttt{\{degradation\_label\}}. \\
8 & Please proceed to the second \texttt{\{audio\_type\}} and listen carefully to the \texttt{\{degradation\_label\}}. \\
9 & Next, listen carefully to the second comparison clip. It \texttt{\{be\_verb\}} \texttt{\{degradation\_label\}}. \\
10 & Now listen to the second audio example and pay attention to the \texttt{\{degradation\_label\}}. \\
11 & Please listen to the second sample carefully and note the perceptual characteristics of the \texttt{\{degradation\_label\}}. \\
12 & The second clip serves as the next comparison recording. It \texttt{\{be\_verb\}} \texttt{\{degradation\_label\}}. Listen closely. \\
13 & Use the second \texttt{\{audio\_type\}} as the next comparison sample and listen carefully to the \texttt{\{degradation\_label\}}. \\
14 & Please listen to the second recording carefully and note the perceived effect of the \texttt{\{degradation\_label\}}. \\
15 & Now listen to the second clip and remember the perceptual effect of the \texttt{\{degradation\_label\}} for later comparison. \\

\midrule
\multicolumn{2}{c}{\textbf{Round~3 Introduction Prompt}} \\
\midrule
1 & Now listen to the third \texttt{\{audio\_type\}}. \\
2 & Please now listen carefully to the third \texttt{\{audio\_type\}}. \\
3 & Next, listen to the third \texttt{\{audio\_type\}}. \\
4 & Now listen to the third comparison \texttt{\{audio\_type\}}. \\
5 & Please listen to the third clip carefully. \\
6 & Continue by listening to the third \texttt{\{audio\_type\}}. \\
7 & Now listen carefully to the third recording. \\
8 & Please proceed to the third \texttt{\{audio\_type\}} and listen carefully. \\
9 & Next, listen carefully to the third comparison clip. \\
10 & Now listen to the third audio example. \\

\midrule
\multicolumn{2}{c}{\textbf{Round~3 Option Format}} \\
\midrule
1 & Please choose the best answer from the following options: A. \texttt{\{A\}}, B. \texttt{\{B\}}, C. \texttt{\{C\}}, D. \texttt{\{D\}}. \\

\midrule
\multicolumn{2}{c}{\textbf{Answer Constraints}} \\
\midrule
1 & Answer with only A, B, C, or D. \\
2 & Respond with only A, B, C, or D. \\
3 & Your answer must be A, B, C, or D only. \\
4 & Please reply using only A, B, C, or D. \\
5 & Output only a single letter: A, B, C, or D. \\
6 & Return only A, B, C, or D as your answer. \\

\bottomrule
\end{tabular}
\vspace{-2mm}
\caption{
DSR general prompt templates.
For each DSR instance, \texttt{\{audio\_type\}} and \texttt{\{degradation\_label\}} are determined by the corresponding audio, while the remaining prompt components are sampled per question.
}
\label{tab:dsr_general_templates}
\end{table*}

\begin{table*}[t]
\centering
\scriptsize
\setlength{\tabcolsep}{4pt}
\renewcommand{\arraystretch}{0.934}
\begin{tabular}{
>{\centering\arraybackslash}p{0.02\textwidth}
p{0.4625\textwidth}
p{0.4625\textwidth}
}
\toprule
\textbf{ID} 
& \textbf{Ranking Directions: \texttt{\{low\_word\}} $\boldsymbol{\rightarrow}$ \texttt{\{high\_word\}}} 
& \textbf{Ranking Directions: \texttt{\{high\_word\}} $\boldsymbol{\rightarrow}$ \texttt{\{low\_word\}}} \\
\midrule
1 & lower severity $\rightarrow$ higher severity & higher severity $\rightarrow$ lower severity\\
2 & lower intensity $\rightarrow$ greater intensity & greater intensity $\rightarrow$ lower intensity \\
3 & lower degree $\rightarrow$ higher degree & higher degree $\rightarrow$ lower degree \\
4 & less severe degradation $\rightarrow$ less severe degradation & more severe degradation $\rightarrow$ less severe degradation \\
5 & lower artifact level $\rightarrow$ higher artifact level & higher artifact level $\rightarrow$ lower artifact level \\

\midrule
\multicolumn{3}{c}{\textbf{Round~3 Text Prompt (Ambient Noise)}} \\
\midrule
1 & All three clips contain ambient noise. Rank them from \texttt{\{low\_word\}} to \texttt{\{high\_word\}}.
  & All three clips contain ambient noise. Rank them from \texttt{\{high\_word\}} to \texttt{\{low\_word\}}. \\
2 & Rank the three clips based on ambient noise from \texttt{\{low\_word\}} to \texttt{\{high\_word\}}.
  & Rank the three clips based on ambient noise from \texttt{\{high\_word\}} to \texttt{\{low\_word\}}. \\
3 & Order the three clips from less audible ambient noise to more audible ambient noise.
  & Order the three clips from more audible ambient noise to less audible ambient noise. \\
4 & Determine which ordering reflects ambient noise becoming progressively more noticeable.
  & Determine which ordering reflects ambient noise becoming progressively less noticeable. \\

\midrule
\multicolumn{3}{c}{\textbf{Low-pass Filtering}} \\
\midrule
1 & All three clips are processed with low-pass filtering. Rank them from \texttt{\{low\_word\}} to \texttt{\{high\_word\}}.
  & All three clips are processed with low-pass filtering. Rank them from \texttt{\{high\_word\}} to \texttt{\{low\_word\}}. \\
2 & Rank the three clips based on low-pass filtering from \texttt{\{low\_word\}} to \texttt{\{high\_word\}}.
  & Rank the three clips based on low-pass filtering from \texttt{\{high\_word\}} to \texttt{\{low\_word\}}. \\
3 & Order the three clips from less muffled to more muffled.
  & Order the three clips from more muffled to less muffled. \\
4 & Determine which ordering reflects increasing loss of high-frequency content.
  & Determine which ordering reflects decreasing loss of high-frequency content. \\

\midrule
\multicolumn{3}{c}{\textbf{Band-pass Filtering}} \\
\midrule
1 & All three clips are processed with band-pass filtering. Rank them from \texttt{\{low\_word\}} to \texttt{\{high\_word\}} in terms of band-limited severity.
  & All three clips are processed with band-pass filtering. Rank them from \texttt{\{high\_word\}} to \texttt{\{low\_word\}} in terms of band-limited severity. \\
2 & All three clips are affected by band-pass filtering. Rank the three clips in terms of band-limited severity from \texttt{\{low\_word\}} to \texttt{\{high\_word\}}.
  & All three clips are affected by band-pass filtering. Rank the three clips in terms of band-limited severity from \texttt{\{high\_word\}} to \texttt{\{low\_word\}}. \\
3 & All three clips contain artifacts caused by band-pass filtering. Order the three clips from less band-limited to more band-limited.
  & All three clips contain artifacts caused by band-pass filtering. Order the three clips from more band-limited to less band-limited. \\
4 & All three clips exhibit band-limiting effects due to band-pass filtering. Determine which ordering reflects increasing band-limiting.
  & All three clips exhibit band-limiting effects due to band-pass filtering. Determine which ordering reflects decreasing band-limiting. \\

\midrule
\multicolumn{3}{c}{\textbf{Reverberation}} \\
\midrule
1 & All three clips are affected by reverberation. Rank them from \texttt{\{low\_word\}} to \texttt{\{high\_word\}}.
  & All three clips are affected by reverberation. Rank them from \texttt{\{high\_word\}} to \texttt{\{low\_word\}}. \\
2 & Rank the three clips based on reverberation from \texttt{\{low\_word\}} to \texttt{\{high\_word\}}.
  & Rank the three clips based on reverberation from \texttt{\{high\_word\}} to \texttt{\{low\_word\}}. \\
3 & Order the three clips from less reverberant to more reverberant.
  & Order the three clips from more reverberant to less reverberant. \\
4 & Determine which ordering reflects increasingly strong reverberant tails.
  & Determine which ordering reflects progressively weaker reverberant tails. \\

\midrule
\multicolumn{3}{c}{\textbf{Echo}} \\
\midrule
1 & All three clips are mixed with echo. Rank them from \texttt{\{low\_word\}} to \texttt{\{high\_word\}}.
  & All three clips are mixed with echo. Rank them from \texttt{\{high\_word\}} to \texttt{\{low\_word\}}. \\
2 & Rank the three clips based on echo from \texttt{\{low\_word\}} to \texttt{\{high\_word\}}.
  & Rank the three clips based on echo from \texttt{\{high\_word\}} to \texttt{\{low\_word\}}. \\
3 & Order the three clips from less prominent echo to more prominent echo.
  & Order the three clips from more prominent echo to less prominent echo. \\
4 & Determine which ordering reflects increasingly noticeable delayed repetitions.
  & Determine which ordering reflects progressively weaker delayed repetitions. \\

\midrule
\multicolumn{3}{c}{\textbf{MP3 Compression}} \\
\midrule
1 & All three clips are compressed using MP3. Rank them from \texttt{\{low\_word\}} to \texttt{\{high\_word\}}.
  & All three clips are compressed using MP3. Rank them from \texttt{\{high\_word\}} to \texttt{\{low\_word\}}. \\
2 & Rank the three clips based on MP3 compression artifacts from \texttt{\{low\_word\}} to \texttt{\{high\_word\}}.
  & Rank the three clips based on MP3 compression artifacts from \texttt{\{high\_word\}} to \texttt{\{low\_word\}}. \\
3 & Order the three clips from less distorted by MP3 compression to more distorted by MP3 compression.
  & Order the three clips from more distorted by MP3 compression to less distorted by MP3 compression. \\
4 & Determine which ordering reflects increasingly audible codec artifacts.
  & Determine which ordering reflects decreasing audibility of codec artifacts. \\

\midrule
\multicolumn{3}{c}{\textbf{Neural Vocoder Artifacts}} \\
\midrule
1 & All three clips contain neural vocoder artifacts. Rank them from \texttt{\{low\_word\}} to \texttt{\{high\_word\}}.
  & All three clips contain neural vocoder artifacts. Rank them from \texttt{\{high\_word\}} to \texttt{\{low\_word\}}. \\
2 & Rank the three clips based on neural vocoder artifacts from \texttt{\{low\_word\}} to \texttt{\{high\_word\}}.
  & Rank the three clips based on neural vocoder artifacts from \texttt{\{high\_word\}} to \texttt{\{low\_word\}}. \\
3 & Order the three clips from fewer vocoder artifacts to more vocoder artifacts.
  & Order the three clips from more vocoder artifacts to fewer vocoder artifacts. \\
4 & Determine which ordering reflects increasingly noticeable vocoder reconstruction distortion.
  & Determine which ordering reflects decreasing vocoder reconstruction distortion. \\

\midrule
\multicolumn{3}{c}{\textbf{DSP Denoising Artifacts}} \\
\midrule
1 & All three clips contain DSP denoising artifacts. Rank them from \texttt{\{low\_word\}} oversuppression to \texttt{\{high\_word\}} oversuppression.
  & All three clips contain DSP denoising artifacts. Rank them from \texttt{\{high\_word\}} oversuppression to \texttt{\{low\_word\}} oversuppression. \\
2 & Rank the three clips based on DSP denoising artifacts from \texttt{\{low\_word\}} oversuppression to \texttt{\{high\_word\}} oversuppression.
  & Rank the three clips based on DSP denoising artifacts from \texttt{\{high\_word\}} oversuppression to \texttt{\{low\_word\}} oversuppression. \\
3 & Order the three clips from less oversuppression to more oversuppression.
  & Order the three clips from more oversuppression to less oversuppression. \\
4 & Determine which ordering reflects increasing oversuppression artifacts caused by denoising algorithm processing.
  & Determine which ordering reflects decreasing oversuppression artifacts caused by denoising algorithm processing. \\

\midrule
\multicolumn{3}{c}{\textbf{Packet Loss}} \\
\midrule
1 & All three clips are affected by packet loss. Rank them from \texttt{\{low\_word\}} to \texttt{\{high\_word\}}.
  & All three clips are affected by packet loss. Rank them from \texttt{\{high\_word\}} to \texttt{\{low\_word\}}. \\
2 & Rank the three clips based on packet loss from \texttt{\{low\_word\}} to \texttt{\{high\_word\}}.
  & Rank the three clips based on packet loss from \texttt{\{high\_word\}} to \texttt{\{low\_word\}}. \\
3 & Order the three clips from fewer audible glitches and dropouts to more audible glitches and dropouts.
  & Order the three clips from more audible glitches and dropouts to fewer audible glitches and dropouts. \\
4 & Determine which ordering reflects increasingly severe temporal discontinuities.
  & Determine which ordering reflects decreasing temporal discontinuities. \\

\bottomrule
\end{tabular}
\vspace{-2mm}
\caption{
DSR Round~3 ranking phrases and degradation-specific text prompt templates.
}
\label{tab:dsr_round3_templates}
\end{table*}

\end{document}